\newif\ifsingle
\singletrue 

\ifsingle
	\documentclass[12pt]{article}
\else	
	\documentclass[10pt,twocolumn,journal]{IEEEtran}
\fi

\ifsingle
	\def \picwidth {10cm} 
\else 
	\def \picwidth {0.8\columnwidth}
\fi

\newif\ifnoappc
\noappctrue

\usepackage[top=1in, bottom=0.75in, left=0.75in, right=0.75in]{geometry}
\usepackage{graphicx}
\usepackage{amssymb}
\usepackage{epsfig}
\usepackage{subfigure}
\usepackage{epstopdf}
\usepackage{algorithm}
\usepackage{algorithmic}
\usepackage{amsmath}
\usepackage{amsfonts}
\usepackage{dsfont}
\usepackage{url}
\usepackage{chemarr}
\usepackage{chemarrow}
\usepackage{bbold}
\usepackage[version=3]{mhchem}
\usepackage{mathabx}
\usepackage{xcolor}
\usepackage{tabularx}

\begin{document} 

\title{A Markovian Approach to the Optimal Demodulation of Diffusion-based Molecular Communication Networks} 
\author{Chun Tung Chou \\
School of Computer Science and Engineering, \\ University of New South Wales, \\ Sydney, Australia. 
E-mail: ctchou@cse.unsw.edu.au}

%

\maketitle

\begin{abstract} 
In a diffusion-based molecular communication network, transmitters and receivers communicate by using signalling molecules (or ligands) in a fluid medium. This paper assumes that the transmitter uses different chemical reactions to generate different emission patterns of signalling molecules to represent different transmission symbols, and the receiver consists of receptors. When the signalling molecules arrive at the receiver, they may react with the receptors to form ligand-receptor complexes. Our goal is to study the demodulation in this setup assuming that the transmitter and receiver are synchronised. We derive an optimal demodulator using the continuous history of the number of complexes at the receiver as the input to the demodulator. We do that by first deriving a communication model which includes the chemical reactions in the transmitter, diffusion in the transmission medium and the ligand-receptor process in the receiver. This model, which takes the form of a continuous-time Markov process, captures the noise in the receiver signal due to the stochastic nature of chemical reactions and diffusion. We then adopt a maximum a posteriori framework and use Bayesian filtering to derive the optimal demodulator. We use numerical examples to illustrate the properties of this optimal demodulator. 
\end{abstract} 


\section{Introduction} 
\label{sec:intro} 
Molecular communication is a promising approach to realise communications among nano-scale devices \cite{Akyildiz:2008vt,Hiyama:2010jf,Nakano:2012dv,Nakano:2014fq}. There are many possible applications with these networks of nano-devices, for example, in-body sensor networks for health monitoring and therapy \cite{Atakan:2012ej,Nakano:2012dv}. This paper considers diffusion-based molecular communication networks. 

In a diffusion-based molecular communication network, transmitters and receivers communicate by using signalling molecules or ligands. The transmitter uses different time-varying functions of concentration of signalling molecules (or emission patterns) to represent different transmission symbols. The signalling molecules diffuse freely in the medium. When signalling molecules reach the receiver, they react with chemical species in the receiver to produce output molecules. The counts of output molecules over time is the receiver output signal which the receiver uses to decode the transmitted symbols. 

Two components in diffusion-based molecular communication system are modulation and demodulation. A number of different modulation schemes have been considered in the literature. For example, \cite{Atakan:2010bj,Mahfuz:2011te} consider Concentration Shift Keying (CSK) where different concentrations of signalling molecules are used by the transmitter to represent different transmission symbols. Other modulation techniques that have been proposed include Molecule Shift Keying (MSK) \cite{ShahMohammadian:2012iu,Kuran:2011tg}, Pulse Position Modulation (PPM) \cite{Kadloor:ua}, 
Amplitude Shift Keying (ASK)\cite{Mahdavifar15}, Frequency Shift Keying (FSK) \cite{Chou:2012ug}, and token communication \cite{Rose:hn}. This paper assumes that the transmitter uses different chemical reactions to generate the emission patterns of different transmission symbols. The motivation to use this type of modulation mechanism is that chemical reactions are a natural way to produce signalling molecules, e.g. the papers \cite{Ni:2010de,Jovic:2010kb} study a number of molecular circuits (which are sets of chemical reactions) that can produce oscillating signals, and the paper \cite{Tostevin:2010bo} discusses a number of signalling mechanisms in living cells. 


We assume the receiver consists of receptors. When the signalling molecules (ligands) reach the receiver, they can react with the receptors to form ligand-receptor complexes (which are the output molecules in this paper). We consider the problem of using the continuous-time history of the number of complexes for demodulation assuming that the transmitter and receiver are synchronised. The ligand-receptor complex signal is a stochastic process with three sources of noise because the chemical reactions at the transmitter, the diffusion of signalling molecules and the ligand-receptor binding process are all stochastic. We derive a continuous-time Markov process (CTMP) which models the chemical reactions at the transmitter, the diffusion in the medium and the ligand-receptor binding process. By using this model and the theory of Bayesian filtering, we derive the maximum a posteriori (MAP) demodulator using the continuous-time history of the number of complexes as the input. 

This paper makes two key contributions: (1) We propose to use a CTMP to model a molecular communication network with chemical reactions at the transmitter, a diffusive propagation propagation medium and receptors at the receiver. The CTMP captures all three sources of noise in the communication network. (2) We derive a closed-form expression for the MAP demodulation filter using the proposed CTMP. The closed-form expression gives insight into the important elements needed for optimal demodulation, these are the timings at which the receptor bindings occur, the number of unbound receptors and the mean concentration of signalling molecules around the receptors. 

The rest of the paper is organised as follows. Section \ref{sec:related} discusses related work. Section \ref{sec:model} presents the system assumptions, as well as a mathematical model from the transmitter to the ligand-receptor complex signal based on CTMP. We derive the MAP demodulator in Section \ref{sec:MAP} and illustrate its numerical properties in Section \ref{sec:eval}. Finally, Section \ref{sec:con} concludes the paper.

\section{Related work} 
\label{sec:related}
There is a growing interest to understand molecular communication from the communication engineering point of view. For recent surveys of the field, see \cite{Akyildiz:2008vt,Hiyama:2010jf,Nakano:2012dv,Nakano:2014fq,Akyildiz:hy}. We divide the discussion under these headings: transmitters, receivers, models and others. 

{\bf Transmitters.} A number of different types of transmission signals have been considered in the molecular communication literature. The papers \cite{ShahMohammadian:2013jm,Mahfuz:2011te} assume that the transmitter releases the signalling molecules in a burst which can be modelled as either an impulse or a pulse with a finite duration. A recent work in \cite{Mosayebi:2014dm} assumes that the transmitter releases the molecules according to a Poisson process. In this paper, we instead assume that the transmitter uses different sets of chemical reactions to generate different transmission symbols and we use CTMP to model these transmission symbols. Since a Poisson process can also be modelled by a CTMP, the transmission process in this paper is more general than that of \cite{Mosayebi:2014dm}. Our CTMP model can also deal with an impulsive input by using an appropriate initial condition for the CTMP. The use of CTMP as an end-to-end model --- which includes the transmitter, the medium and the receiver --- does not appear to have been used before. 


{\bf Receivers. } 
Demodulation methods for diffusion-based molecular communication have been studied  in \cite{Noel:2014hu,Mahfuz:2014vs}. Both papers also use the MAP framework with discrete-time samples of the number of output molecules as the input to the demodulator. Instead, in this paper, we consider demodulation using continuous-time history of the number of complexes. 

The demodulation from ligand-receptor signal has also been considered in \cite{ShahMohammadian:2013jm}. The key difference is that \cite{ShahMohammadian:2013jm} uses a linear approximation of the ligand-receptor process while we use a non-linear reaction rate. 

The capacity of molecular communications based on ligand-receptor binding has been studied in \cite{Einolghozati:2011ge,Einolghozati:2011cj} assuming discrete samples of the number of complexes are available. A recent work \cite{Thomas:2014tf} considers the capacity of such systems in the continuous-time limit. Instead of focusing on the capacity, our work focuses on demodulation. 

Receiver design is an important topic in molecular communication and has been studied in many papers, some examples are \cite{Noel:2014fv,Noel:2014hu,Kilinc:2013by,Mahfuz:2014vs,Meng:2014hh}. These papers either use one sample or a number of discrete samples on the count of a specific molecule to compute the likelihood of observing a certain input symbols. This paper takes a different approach and uses continuous-time signals. 

Another approach of receiver design for molecular communication is to derive molecular circuits that can be used for decoding. An attempt is made in \cite{Chou:2012ug} to design a molecular circuit that can decode frequency-modulated signals. However, the work does not take diffusion and reaction noise into consideration. A recent work in \cite{Pierobon:2014iu} analyses end-to-end molecular communication biological circuits from linear time-invariant system point of view. The work in \cite{Chou:2014jca} compares the information theoretic capacity of a number of different types of linear molecular circuits. This paper differs from the previous work in that it uses a non-linear ligand-receptor binding model. 

The noise property of ligand-receptor for molecular communication has been characterised in \cite{Pierobon:2011ve}. The case for non-linear ligand-receptor binding does not appear to have an analytical solution and \cite{Pierobon:2011ve} derives an approximate characterisation using a linear reaction rate assuming that the number of signalling molecules around the receptor is large. This paper uses a non-linear ligand-receptor binding model and no approximation is used in solving the filtering problem. 


{\bf Models.} This paper uses the Reaction Diffusion Master Equation (RDME) \cite{Gillespie:2013kk} framework to model the reactions and diffusion in the molecular communication networks. RDME assumes that time is continuous while the diffusion medium is discretised into voxels. This results in a CTMP with finite number of (discrete) states. RDME has been used to model stochastic dynamics of cells in the biology literature \cite{Lawson:2013fe}.  An attraction of RDME is that it has the Markov property which means that one can leverage the rich theory behind Markov process. 

The author of this paper has previously used an extension of the RDME model, called the RDME with exogenous input (RDMEX) model, to study molecular communication networks in \cite{Chou:rdmex_tnb,Chou:rdmex_nc,Chou:tnano_impact,Chou:tnano_analog}. The RDMEX assumes that the times at which the transmitter emits signalling molecules are deterministic. This results in a stochastic process which is piecewise Markov or the Markov property only holds in between two consecutive emissions by the transmitter. In this paper, we assume the transmitter uses chemical reactions to generate the signalling molecules. Therefore, the emission timings are not deterministic but are governed by a stochastic process. 

In this paper, we assume that the propagation medium is discretised in the voxels. An alternative modelling paradigm that has been used in a number of molecular communication network papers \cite{Mahfuz:2011te,ShahMohammadian:2013jm,Mosayebi:2014dm} is that the transmitter or receiver has a non-zero spatial dimension (commonly modelled by a sphere) while the propagation medium is assumed to be continuous. (Note that though \cite{Mosayebi:2014dm} does not explicitly state the dimension of the receiver, one can infer from the fact that the receiver must have a non-zero dimension because it has a non-zero probability of receiving the signalling molecules.) We believe the technique in this paper can be adapted to this alternative modelling paradigm and we do not expect this alternative modelling paradigm will change the results in this paper; we will explain this in Section \ref{sec:map:css}. 

There is a rich literature in the modelling of biological systems discussing the difference between: (1) The particle approach which has a continuous state space because the state of a particle is its position; and (2) The mesoscopic approach (the approach in this paper) which discretises the medium into discrete voxels and consider the number of molecules in the voxels as the state. The first approach is more accurate but the computation burden can be high \cite{Burrage:2011te}, while the second approach is accurate for appropriate discretisation \cite{Hellander:2011ub,Gillespie:2013kk}. There are also hybrid approaches too. An overview of various modelling and simulation approaches can be found in \cite{Burrage:2011te}. 

{\bf Others:} The results of this paper may also be of interest to biologists who are interested to understand how living cells can distinguish between different concentration levels \cite{Siggia:2013dd,Kobayashi:2011dh}. The result of this paper can be viewed as a generalisation of \cite{Siggia:2013dd} which studies how cells can distinguish between two constant levels of ligand concentration.

\section{End-to-end communication models} 
\label{sec:model} 
\begin{figure}
\begin{center}
\includegraphics[page=2,trim=0cm 0cm 0cm 0cm ,clip=true, width=\picwidth]{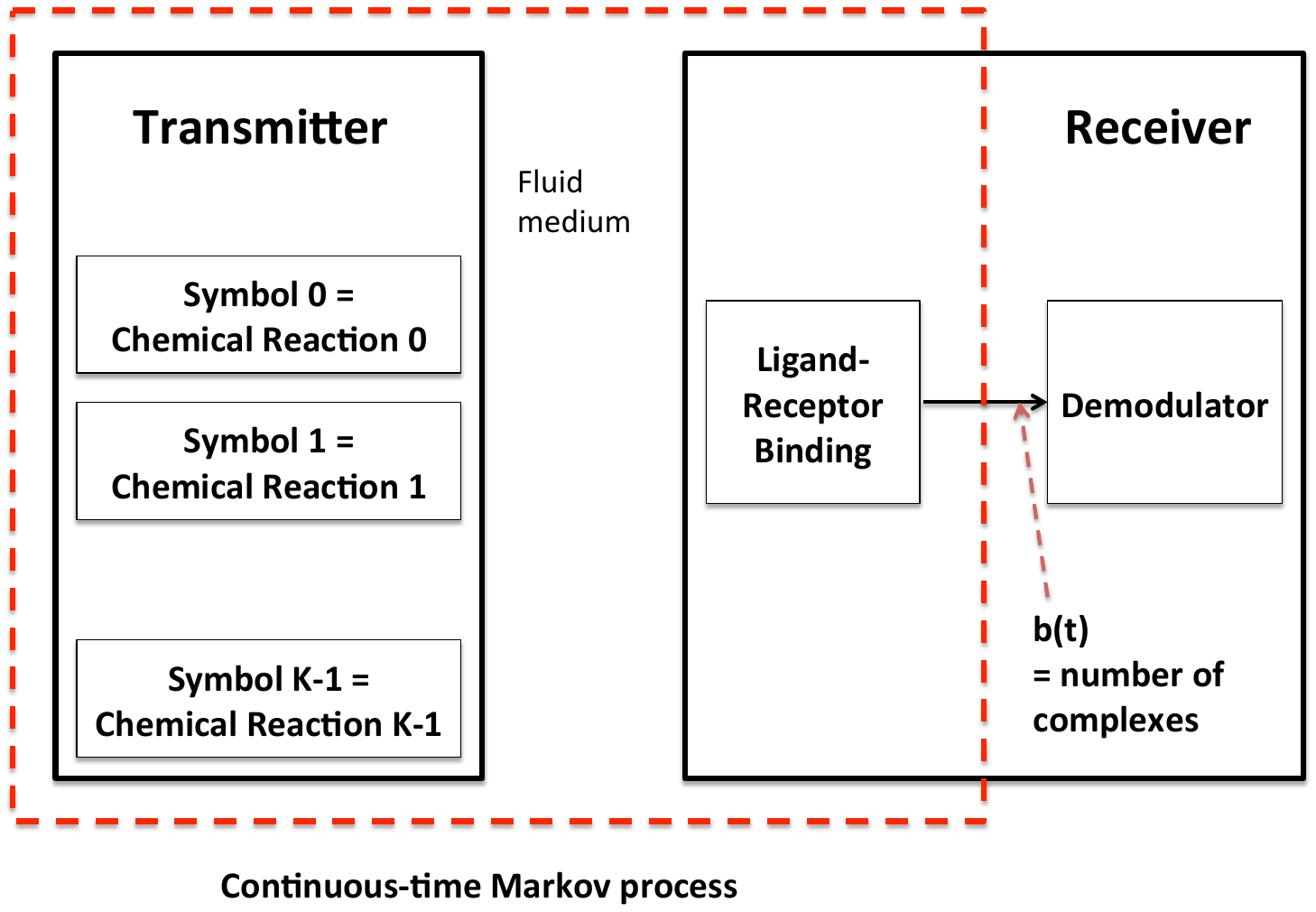}
\caption{An overview of the system considered in this paper.}
\label{fig:overall}
\end{center}
\end{figure}

This paper considers diffusion-based molecular communication with one transmitter and one receiver in a fluid medium. Figure \ref{fig:overall} gives an overview of the setup considered in this paper. The transmitter uses different chemical reactions to generate the emission patterns of different transmission symbols. The transmitter acts as the source and emitter of signalling molecules. The signalling molecules diffuses in the fluid medium. The front-end of the receiver consists of a ligand-receptor binding process and the back-end consists of the demodulator with the number of complexes as its input. 

In this section, we first describe the system assumptions in Section \ref{sec:model_basic}. We then present, in Section \ref{sec:e2e:g}, an end-to-end model which includes the transmitter, the transmission medium and the ligand-receptor binding process in the receiver, see the dashed box in Figure \ref{fig:overall}. The end-to-end model is a CTMP which includes chemical reactions in the transmitter, diffusion in the medium and the ligand-receptor binding process in the receiver. 


\subsection{Model assumptions}
\label{sec:model_basic} 
We assume that the medium (or space) is discretised into voxels while time is continuous. This modelling framework results in a RDME \cite{Gillespie:2013kk,Gardiner,vanKampen}, which is a CTMP commonly used to model systems with both diffusion and reactions. In addition, we assume the communication uses only one type of signalling molecule (or ligand) denoted by $S$. We divide the description of our model into three parts: transmission medium, transmitter and receiver. We begin with the transmission medium. Table \ref{tab:notation} summaries the frequently used notation and chemical symbols. 

\begin{table}
\begin{tabularx}{\textwidth}{|c|X|} \hline 
Symbol & Meaning \\ \hline 
$W$     &  Dimension of one side of a voxel \\
$D$      & Diffusion constant \\ 
$d$      & Diffusion rate between neighbouring voxels \\ 
$N_v$  & Total number of voxels \\
$M$ & Total number of receptors \\
$\tilde{\lambda}$ & Reaction rate constant for the binding reaction \\
$\lambda$ & $\lambda = \frac{\tilde{\lambda}}{W^3}$ \\
$\mu$ & Reaction rate constant for the unbinding reaction \\ 
$s$ & A transmission symbol \\
$b(t)$  & Number of complexes at time $t$ \\
$n_i(t)$ & Number of signalling molecules in voxel $i$ at time $t$\\ 
$N(t)$ & Equation \eqref{eqn:state}. A vector containing the number of signalling molecules in each voxel, the counts of intermediate
  chemical species in the transmitter and the cumulative count of the number of molecules that have left the system \\
$\sigma_s(t)$ & The mean number of signalling molecules in the receiver voxel at time $t$ if the transmitter sends symbol $s$\\
$U(t)$ & The cumulative number of times the receptors have switched from the unbound to bound state at time $t$\\
$E$      & An unbound receptor \\
$S$      & A signalling molecule \\
$C$      & A complex  \\
\hline 
\end{tabularx}
\caption{Notation and chemical symbols}
\label{tab:notation}
\end{table}

\subsubsection{Transmission medium} 
\begin{figure}
\begin{center}
\includegraphics[page=1,trim=0cm 0cm 0cm 0cm ,clip=true, width=\picwidth]{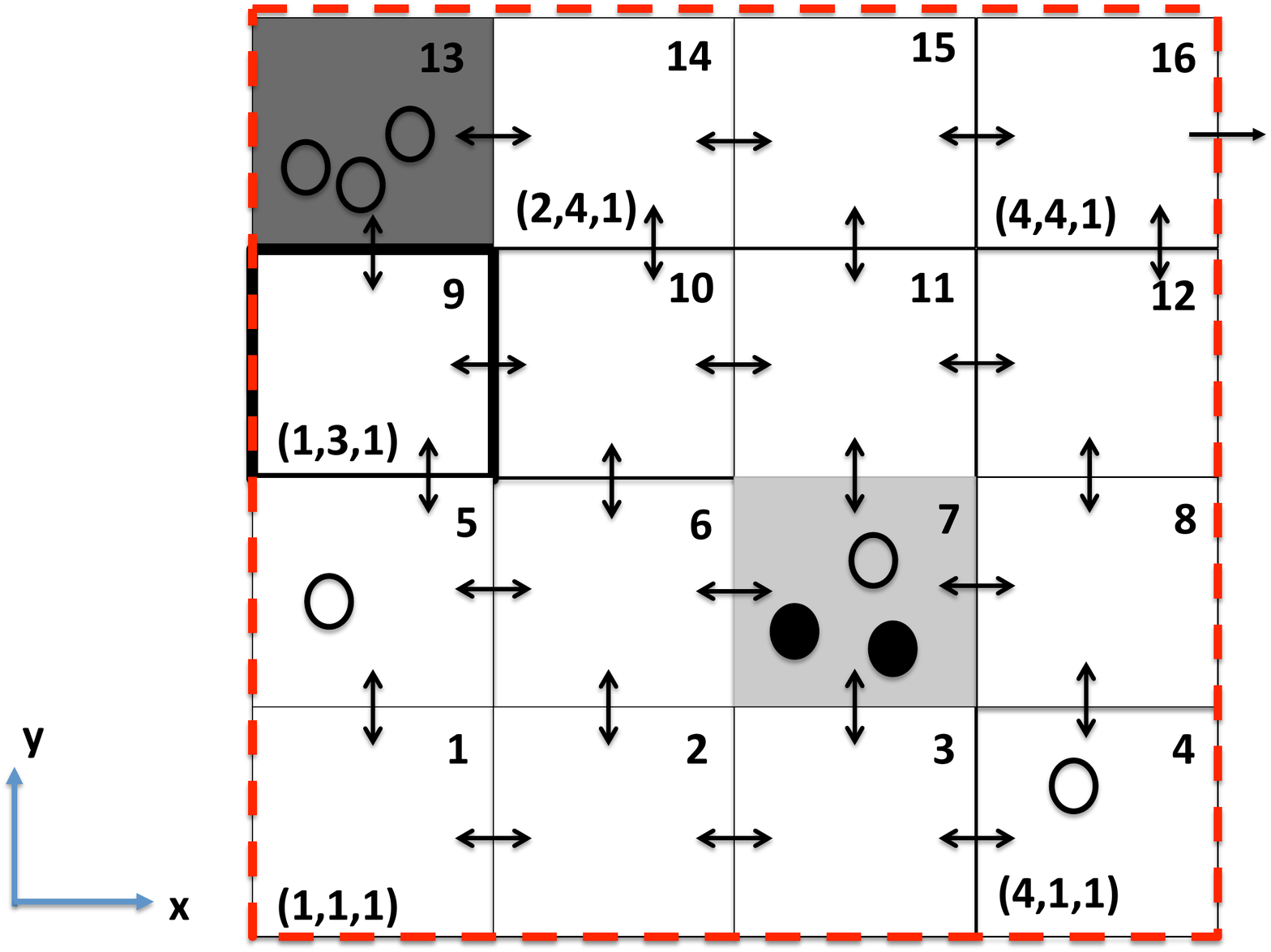}
\caption{A model of molecular communication network. The volume is divided into voxels. The indices of the voxels are given in the top right hand corner. Unfilled circles are signalling molecules. Filled circles are receptors. The dashed lines show the boundary of the medium.}
\label{fig:model}
\end{center}
\end{figure}

We model the transmission medium as a three dimensional (3-D) space and partition the space into cubic {\sl voxels} of volume $W^3$. Figure \ref{fig:model} shows an example of a medium which has a dimension of 4 voxels along both the $x$ and $y$-directions, and 1 voxel in the $z$-direction. (Note that Figure \ref{fig:model} should be viewed as a projection onto the $x-y$ plane.) In general, we assume the medium to have $N_x$, $N_y$ and $N_z$ voxels in the $x$, $y$ and $z$ directions where $N_x, N_y$ and $N_z$ are positive integers. In Figure \ref{fig:model}, $N_x = N_y = 4$ and $N_z = 1$. We also use $N_v = N_x N_y N_z$ to denote the total number of voxels. 

We refer to a voxel by a triple $(x,y,z)$ where $x$, $y$ and $z$ are integers or by a single index $\xi \in [1,N_v]$. Figure \ref{fig:model} shows the triples for some of the voxels. The single index $\xi$ is calculated from the triple $(x,y,z)$ by using $\xi(x,y,z) = x + N_x (y-1) + N_x N_y (z-1)$. The single indices for voxels are shown in the top right-hand corner of the voxels in Figure \ref{fig:model}.


Diffusion is modelled by molecules moving from one voxel to a neighbouring voxel. For examples, in Figure \ref{fig:model}, molecules can diffuse from Voxel 1 to Voxels 2 or 5, from Voxel 2 to Voxels 1, 3 and 6, and so on. The diffusion of molecules between neighbouring voxels is indicated by the two-way arrows in Figure \ref{fig:model}. 

We assume that the signalling molecule $S$ is the only diffusible chemical species in our model and the diffusion coefficient for $S$ is $D$. This means the signalling molecules diffuse from one voxel to a neighbouring voxel at a mean rate of $d$ where $d = \frac{D}{W^2}$. In other words, within an infinitesimal time $\Delta t$, the probability that a signalling molecule diffuses to a neighbouring voxel is $d \; \Delta t$. Note that the expression for $d$ can be derived from spatially discretising the diffusion equation into regular cubic voxels of volume $W^3$, see \cite[p.341]{Gardiner} or \cite[Section 3]{Erban:2007we}.

The dashed lines in Figure \ref{fig:model} indicate the boundary of our transmission medium. Many different boundary conditions are used by engineers and physicists to model what happens when a molecule reaches the boundary of a medium. Two typical boundary conditions are absorbing and reflecting boundaries \cite{Gardiner}. An absorbing boundary means that a molecule can leave the transmission medium and once the molecule has left, it will not return to the medium. For example, in Figure \ref{fig:model}, we allow molecules to leave the medium via one surface of Voxel 16 as indicated by the one-way arrow. Mathematically, this is modelled by a rate of leaving the medium, similar to that of modelling the diffusion between the voxels. A reflecting boundary means that a molecule cannot leave the medium, i.e. a molecule hitting a reflecting boundary will stay in the voxel. Our model can capture these boundary conditions. 

It has been shown in \cite{Hellander:2011ub,Gillespie:2013kk} that in order for RDME to produce physically meaningful results, the voxel dimension $W$ cannot be too small and in order for RDME to reduce the discretisation error, $W$ cannot be too large. In this paper, we assume that $W$ comes from a valid range. The choice of $W$ is beyond the scope of the paper and the reader can refer to \cite{Hellander:2011ub,Gillespie:2013kk} for further discussion. 

For simplicity, we assume that the medium is homogeneous with a constant diffusion coefficient $D$. It is straightforward to extend the framework to cover inhomogeneous medium \cite{Chou:tnano_impact}. It is also possible to use non-cubic voxels, see \cite{Isaacson:2007ui,Engblom:2009tr}. 

\subsubsection{Transmitter} 
\label{sec:model:transmitter}
We assume the transmitter occupies one voxel. However, it is straightforward to generalise to the case where a transmitter occupies multiple voxels. We limit our consideration to one symbol interval without inter-symbol interference (ISI) at this moment. We will discuss the multiple symbol interval case with ISI in Section \ref{sec:isi}. 



We assume that the transmitter can send $K$ different symbols $s = 0,1,..,K-1$ where each symbol $s$ is characterised by an emission pattern $u_s(t)$. The role of emission pattern in molecular communication is the same as that of transmitted signal in electromagnetic communication. If a transmitter uses a deterministic emission pattern $u_s(t)$ to represent symbol $s$, it means the transmitter emits $u_s(t)$ signalling molecules into the transmitter voxel at time $t$. We use an example to illustrate the meaning of emission pattern. Consider an emission pattern $u_1(t)$ for Symbol 1 where  $u_1(t) = \delta_{t,1.2} + \delta_{t,5.6} + 2 \delta_{t,8.1}$ where $\delta_{i,j}$ denotes the Kronecker delta, which means $\delta_{i,j} = 1$ if and only if $i = j$. The emission pattern $u_1(t)$ means that, for Symbol 1, the transmitter emits one signalling molecule at times 1.2 and 5.6, two signalling molecules at time 8.1 and does not emit any molecules at any other times. 

In this paper, we assume that the emission pattern for each symbol is produced by a set of chemical reactions located in the transmitter voxel. Given $K$ symbols, the transmitter uses $K$ different reactions to generate these symbols, see Figure \ref{fig:overall}. As an example, a class of chemical reactions inside living cells \cite{Tkacik:2011jr} is 
\begin{align}
\cee{
RNA &  ->[\kappa] RNA +  A
} 
\label{cr:rna}
\end{align}
where ribonucleic acid (RNA, which is a molecule commonly found in living cells) produces the chemical species $A$. This class of chemical reactions can be modelled by a Poisson process where molecules of $A$ are produced at a mean rate of $\kappa$ \cite{Shahrezaei:2008bp}. Note that the emission patterns produced by chemical reactions are {\sl not} deterministic, but {\sl stochastic}. The mean emission pattern of this chemical reaction is ${\bf E}[u(t)] = \kappa$. 

Following on from the above example, one can realise Amplitude Shift Keying (ASK) in molecular communication by using different chemical reactions that can produce signalling molecules at different mean rates. For example, if there are four different reactions that can produce signalling molecules at four different mean rates of $\kappa_0$, $\kappa_1$, $\kappa_2$ and $\kappa_3$, then one can use these four different reactions to produce 4 different symbols. Note that it is possible for the four chemical reactions to produce the same emission pattern (or realisation), though with different probabilities. 
 
A standard result in physical chemistry shows that the dynamics of a set of chemical reactions can be modelled by a CTMP \cite{Gillespie:1992tq}. Therefore, we will model the transmitter by a CTMP. Note that, in this paper, we will not specify the sets of chemical reactions used by the transmitter except for simulation because the MAP demodulator does not explicitly depend on the sets of chemical reactions that the transmitter uses.

\subsubsection{Receiver} 
\label{sec:model:rec} 
We assume the receiver occupies one voxel and we use $R$ to denote the index of the voxel at which the receiver is located. In Figure \ref{fig:model}, we assume the receiver is at Voxel 7 (light grey) and hence $R = 7$ for this example. In addition, we assume that the transmitter and receiver voxels are distinct. 

We assume that the receiver has $M$ non-interacting receptors and we use $E$ as the chemical name for an unbound receptor. These receptors are fixed in space and do not diffuse, and they are only found in the receiver voxel. Furthermore, these receptors are assumed to be uniformly distributed in the receiver voxel. 

The receptor $E$ can bind to a signalling molecule $S$ to form a ligand-receptor complex (or complex for short) $C$, which is a molecule formed by combining $E$ and $S$. This is known as ligand-receptor binding in molecular biology literature \cite{Alberts}. The binding reaction can be written as the chemical equation: 
\begin{align}
\cee{
S + E &  ->[\tilde{\lambda}] C
}
\label{cr:on}  
\end{align}
where $\tilde{\lambda}$ is the reaction rate constant. Since the receptors are only found in the receiver voxel, the binding reaction occurs in a volume of $W^3$, which is the volume of a voxel. The rate at which the complexes $C$ is formed is given by the product of $\frac{\lambda}{W^3}$, the number of signalling molecules in the receiver voxel and the number of unbound receptors\footnote{\label{fn:bind} This footnote explains how $\frac{\lambda}{W^3}$ comes about. Consider a chemical reaction where reactants S and E react to form product C. We assume  the reactions are taking place within a volume of $W^3$. Let $c_{\rm S}$, $c_{\rm E}$ and $c_{\rm C}$ be, respectively, the concentration of S, E and C in the volume. The law of mass action says that $\frac{d  c_{\rm C}}{dt} = \tilde{\lambda} \; c_{\rm E} \; c_{\rm S}$. In the case of the CTMP or RDME in this paper, we want to keep track of the number of molecules in a volume (the voxel) instead. Let $n_{\rm S}$, $n_{\rm E}$ and $n_{\rm C}$ be, respectively, the number of S, E and C molecules in the volume. Since concentration and molecule counts are related by $c_{{\rm C}} W^3 = n_{{\rm C}}$ etc, we will, in a mathematically loose way, write $\frac{d  n_{\rm C}}{dt} = \frac{\tilde{\lambda}}{W^3} \; n_{{\rm S}} \; n_{{\rm E}}$. Since $n_{\rm C}$ is a discrete quantity, the derivative $\frac{d  n_{\rm C}}{dt}$ is not defined but we can interpret it as the production rate of molecules $C$. This explains how to convert the law of mass action, which is in terms of concentration, to the rate law used in RDME which is in terms molecular counts. This conversion is also discussed in \cite{Erban:2009us,Erban:2007we}.}. We define $\lambda = \frac{\tilde{\lambda}}{W^3}$ and will use $\lambda$ in the CTMP. Note that this is equivalent to ligand-receptor binding model used in \cite[Section V-B]{Pierobon:2011ve}. 

A ligand-receptor complex $C$ can dissociate into an unbound receptor $E$ and a signalling molecule $S$. This can be represented by the chemical equation
\begin{align}
\cee{
C &  ->[\mu] E + S
}
\label{cr:off}  
\end{align}
where $\mu$ is the reaction rate constant. The rate at which the complexes are dissociating is given by the product of $\mu$ and the number of complexes\footnote{The law of mass action for the dissociation reaction is $\frac{d  c_{\rm C}}{dt} = - \mu \; c_{\rm C}$ where $c_{\rm C}$ is the concentration of the complexes. We can use the same argument in Footnote \ref{fn:bind} to show that the dissociation rate of $C$ is $\mu \; n_{\rm C}$ where $n_{\rm C}$ is the number of complexes. In particular, note that no scaling by volume $W^3$ of the voxel is required. 
}. 

Since a receptor can either be in an unbound state $E$ or in a complex $C$, we have the following conservation relation: the number of unbound receptors plus the number of complexes is equal to the total number of receptors $M$.

\subsection{General end-to-end model} 
\label{sec:e2e:g}
In order to derive the MAP demodulator, we need an end-to-end model which includes the transmitter, the medium and the ligand-binding process, see Figure \ref{fig:overall}. Since chemical reactions (which includes the chemical reactions in the transmitter as well as the ligand-receptor binding process in the receiver) and diffusion can be modelled by CTMP, it is possible to use a CTMP as an end-to-end model. In this section we present a general end-to-end model that includes the transmitter, diffusion and the ligand-receptor process in the receiver. An excellent tutorial introduction to the modelling of chemical reactions and diffusion by using CTMP can be found in \cite{Erban:2007we}. We have also included an example in Appendix \ref{app:e2e_ex}.



%

The aim of the end-to-end model is to determine the properties of the receiver signal from the transmitter signal. The receiver signal in our case is the number of complexes over time. Since the transmitter uses $K$ symbols, the transmitter signal is generated by one of the $K$ sets of chemical reactions. This means that we need $K$ end-to-end models with a model for each of the $K$ symbols or sets of chemical reactions. The principle behind building these $K$ models is identical so without loss of generality, we will assume that the model here is for Symbol 0. We begin with a few definitions. 

Let $n_i(t)$ (where $1 \leq i \leq N_v$) be the number of signalling molecules $S$ in Voxel $i$ at time $t$. In particular, since we have defined $R$ to be the index of the receiver voxel, $n_R(t)$ is the number of signalling molecules in the receiver voxel. We assume the transmitter is a set of chemical reactions which uses $H$ intermediate chemical species $Q_1$, $Q_2$, ... and $Q_H$ and these intermediate species remain in the transmitter voxel. Let $n_{Q_i}(t)$ be the number of chemical species $Q_i$ in the transmitter voxel at time $t$. Molecules may also be degraded or leave the system forever if absorbing boundary condition is used. We use $n_A(t)$ to denote the cumulative number of molecules that have left the system. Note that since $n_i(t), n_{Q_i}(t)$ and $n_A(t)$ are molecular counts, they must belong to the set of non-negative integers ${\mathbb Z}_{\geq 0}$. We define the vector $N(t) \in {\mathbb Z}^{N_v+H+1}_{\geq 0}$ to be:
\begin{align}
N(t) = 
\left[ \begin{array}{ccccccc}
n_{1}(t) & ... & n_{N_v}(t) & n_{Q_1}(t) & ... & n_{Q_H}(t) & n_{A}(t) 
\end{array}
\right]^T
\label{eqn:state}
\end{align} 
where ${}^T$ denotes matrix transpose. 

Let $b(t)$ denote the number of complexes or bound receptors at time $t$ and ${\mathbb Z}_{[0,M]}$ denote the set of integers between $0$ and $M$ inclusively. We require that a valid $b(t)$ must be an element of ${\mathbb Z}_{[0,M]}$. Since there are $M$ receptors, the number of unbound receptor is $M-b(t)$. 


The state of the end-to-end model is the tuple $(N(t),b(t))$. We will now specify the transition probabilities from state $(N(t),b(t))$ to state $(N(t+\Delta t),b(t+\Delta t))$. State transitions can be caused by any one of these events: a chemical reaction in the transmitter, the diffusion of a signalling molecule from a voxel to neighbouring voxel, and the binding or unbinding of a receptor in the receiver. We know from the theory of CTMP that the probability of two events taking place in an infinitesimal duration of $\Delta t$ is of the order of $o(\Delta t^2)$. Intuitively, this means only one event can occur within $\Delta t$. We can divide the transition probabilities from $(N(t),b(t))$ to $(N(t+\Delta t),b(t+\Delta t))$ into 2 groups depending on whether the number of complexes has changed or not in the time interval $(t,t+\Delta t)$. If the number of complexes has changed from time $t$ to $t + \Delta t$, i.e. $b(t+\Delta t) \neq b(t)$, this means either a binding reaction \eqref{cr:on} or a unbinding reaction \eqref{cr:off} has occurred. 

If a binding reaction \eqref{cr:on} has occurred, then the number of signalling molecules in the receiver voxel is decreased by 1 and the number of complexes $b(t)$ is increased by 1. This reaction occurs at a mean rate of $\lambda \; n_R(t) \; (M-b(t))$. We use $\mathbb{1}_i$ to denote the standard basis vector with a `1' at the $i$-th position. We can write the state transition probability of the receptor binding reaction \eqref{cr:on} as: 
\ifsingle
\begin{align}
{\bf P}[N(t+\Delta t) = N(t)-\mathbb{1}_R, b(t+\Delta t) = b(t) + 1 | N(t), b(t)]  
 = \lambda \; n_R(t) \; (M-b(t)) \; \Delta t
\label{eqn:tp:r1}
\end{align}
\else
\begin{align}
& {\bf P}[N(t+\Delta t) = N(t)-\mathbb{1}_R, b(t+\Delta t) = b(t) + 1 | N(t), b(t)]  \nonumber \\ 
& = \lambda \; n_R(t) \; (M-b(t)) \; \Delta t
\label{eqn:tp:r1}
\end{align}
\fi

Recalling that $R$ is the index of the receiver voxel and $n_R(t)$ is the $R$-th element of $N(t)$ in \eqref{eqn:state}, the expression $N(t+\Delta t) = N(t)-\mathbb{1}_R$ is equivalent to $n_R(t+\Delta t) = n_R(t) - 1$, which means the number of signalling molecules in the receiver voxel has decreased by 1. Similarly, the expression $b(t+\Delta t) = b(t) + 1$ says the number of complexes has increased by 1. The right-hand side (RHS) of \eqref{eqn:tp:r1} is the transition probability and is given by the product of mean reaction rate and $\Delta t$. 
 
Similarly, the transition probability of the unbinding reaction is given by: 
\ifsingle
\begin{align} 
{\bf P}[N(t+\Delta t) = N(t)+\mathbb{1}_R, b(t+\Delta t) = b(t) - 1 | N(t), b(t)]  = \mu \; b(t) \; \Delta t
\label{eqn:tp:r2} 
\end{align}
\else
\begin{align} 
& {\bf P}[N(t+\Delta t) = N(t)+\mathbb{1}_R, b(t+\Delta t) = b(t) - 1 | N(t), b(t)] \nonumber \\ 
& = \mu \; b(t) \; \Delta t
\label{eqn:tp:r2} 
\end{align}
\fi
where RHS of \eqref{eqn:tp:r2} is the transition probability. 

We now specify the second group of transition probabilities with $b(t+\Delta t) = b(t)$. These transitions are caused by either a reaction in the transmitter or diffusion of signalling molecules between neighbouring voxels. Let $\eta_i, \eta_j \in  {\mathbb Z}^{N_v+H+1}_{\geq 0}$ be two valid $N(t)$ vectors; let also $\beta \in {\mathbb Z}_{[0,M]}$. For $\eta_i \neq \eta_j$, we write 
\ifsingle
\begin{align}
 {\bf P}[N(t+\Delta t) = \eta_i, b(t+\Delta t) = \beta | N(t) = \eta_j, b(t) = \beta]  
 = d_{ij} \; \Delta t 
\label{eqn:tp:eta}
\end{align}
\else
\begin{align}
& {\bf P}[N(t+\Delta t) = \eta_i, b(t+\Delta t) = \beta | N(t) = \eta_j, b(t) = \beta] \nonumber \\ 
& = d_{ij} \; \Delta t 
\label{eqn:tp:eta}
\end{align}
\fi
where $d_{ij}$ is the transition rate from state $(\eta_j,\beta)$ to state $(\eta_i,\beta)$. Since this transition is due to either a reaction in the transmitter or diffusion, $d_{ij}$ is independent of the number of complexes $\beta$. Depending on the type of transition, the value of $d_{ij}$ can depend on the reaction constants in the transmitter, diffusion rate and some states of $\eta_j$. For example, if the transition from $\eta_j$ to $\eta_i$ is caused by the diffusion of a signalling molecule from Voxel 1 to Voxel 2, we have $\eta_i = \eta_j - {\mathbb 1}_1 + {\mathbb 1}_2$ at a rate of $d \eta_{j,1}$ where $\eta_{j,1}$ is the first element in $\eta_j$ or equivalently the number of signalling molecules in Voxel 1 in state $\eta_j$; so, for this example, $d_{ij} = d \eta_{j,1}$. 
The main advantage of using Equation \eqref{eqn:tp:eta} is that it allows us a cleaner abstraction to solve the Bayesian filtering problem when deriving the MAP demodulator. We also remark that we will not specify the exact expression of $d_{ij}$ because $d_{ij}$'s do not appear explicitly in the demodulator. 

Equations \eqref{eqn:tp:r1}, \eqref{eqn:tp:r2} and \eqref{eqn:tp:eta} specify all the possible state transitions. The probability of no state transition is therefore:
\begin{align}
 & {\bf P}[N(t+\Delta t) = \eta_j,  b(t+\Delta t) = b(t) | N(t) = \eta_j, b(t)] \nonumber \\ 
 & = 
1 - d_{jj} \; \Delta t  - \lambda \; n_R(t) \; (M-b(t)) \; \Delta t - \mu \; b(t) \; \Delta t  
\label{eqn:tp:no}
\end{align}
where 
\begin{align}
d_{jj} = & \sum_{i \neq j} d_{ij} 
\label{eqn:djj} 
\end{align}

We have now specified all the state transition probabilities for Symbol 0. If a different symbol is used, the value of $H$, the dimension of $N(t)$ and the $d_{ij}$ parameters can change. However, the state transition probabilities still can be summarised by Equations of the form \eqref{eqn:tp:r1}, \eqref{eqn:tp:r2} and \eqref{eqn:tp:eta}. In any case, the derivation of the MAP demodulator only requires us to work with one symbol at a time. Hence, we will use Equations \eqref{eqn:tp:r1}-\eqref{eqn:djj} for any transmission symbol. 

\subsection{Discussion} 
Note that the CTMP includes all the three sources of noise in our system, due to chemical reactions in the transmitter, random diffusive movements in the medium and the ligand-receptor binding process at the receiver. Some of these noise components have also been characterised in earlier literature. For example, \cite{Pierobon:2011vr} discusses sampling noise at the transmitter and counting noise at the receiver. One can study these noises using the derived CTMP and let us take sampling noise as an example. For the moment, let us isolate the transmitter voxel from the propagation medium, i.e. we do not allow the signalling molecules to leave or enter the transmitter voxel. In this case, the number of signalling molecules in the transmitter voxel is due entirely to chemical reactions and we use $n_{\rm isolated}(t)$ to denote the number of signalling molecules in the isolated transmitter voxel at time $t$. Now, let us consider the transmitter voxel again but we allow signalling molecules to diffuse in and out of the voxel; we use $n_{\rm connected}(t)$ to denote the number of signalling molecules in the transmitter voxel in this case. It is natural to consider $n_{\rm connected}(t)$ as the transmitter signal because this is the number of signalling molecules in the transmitter voxel. However, $n_{\rm connected}(t)$ can be different from $n_{\rm isolated}(t)$ because signalling molecules can diffuse in and out of the transmitter voxel, which is the cause of sampling noise. 

Equations \eqref{eqn:tp:r1} to \eqref{eqn:tp:no} hold for any valid state $(N(t),b(t))$. If we collect all the transition probability equations for all valid states, then we can form the infinitesimal generator of the CTMP. For a given initial probability distribution of the initial state $(N(0),b(0))$, one can in principle solve the first order ordinary differential equation (ODE) associated with the infinitesimal generator to compute the probability of the number of complexes $b(t)$, or the property of the receiver signal. However, in practice, this ODE is of a very high dimension and it is an active area of research to derive algorithms to solve this ODE efficiently and accurately \cite{Munsky:2006es}. 


\section{The MAP demodulator} 
\label{sec:MAP} 
This section aims to derive the optimal demodulator using the CTMP derived in the previous section. We assume the input to the demodulator is the continuous-time signal $b(t)$ which is the number of complexes at time $t$. There are a number of reasons why we choose to work with the continuous-time signal $b(t)$, rather than its sampled version. First, the signal $b(t)$ may {\bf not} be strictly band limited in the frequency domain. Second, our results show that the optimal demodulator needs to know the time instances at which a receptor is switching from the unbound to bound state. This timing information, which is essentially an impulse, is unfortunately lost by sampling $b(t)$. Third, the solution of the proposed decoding problem can be used to benchmark molecular circuit \cite{Chou:2014jca} based decoders. Since molecular circuits use chemical reactions for computation, they are fundamentally analogue circuits. Fourth, there is an increasing interest in the circuit design community to design low-power analogue signal processing circuits \cite{Barsakcioglu:2014kd}. 

An intermediate step to derive the MAP demodulator is to solve a continuous-time filtering problem. In a filtering problem, one uses all the observations available up till time $t$ to predict the system state at time $t$. For our case, the observations are the number of complexes $b(t)$ in the continuous interval $[0,t]$. (This means the number of observations is infinite because we are considering the continuous-time signal $b(t)$ in a non-zero time interval $[0,t]$.) We use ${\cal B}(t) = \{ b(\tau); 0 \leq \tau \leq t \}$ to denote the section of the signal $b(t)$ in the time interval $[0,t]$. (Note that the $\{ \}$ is not a set notation. Here ${\cal B}(t)$ is a realisation of the number of complexes in $[0,t]$ of the CTMP. This notation is used in some non-linear filtering literature, e.g. \cite{Haykin:1997il}.) We can think of ${\cal B}(t)$ as the history of the number of complexes up till time $t$. The demodulation problem is to use the history ${\cal B}(t)$ to determine which symbol the transmitter has sent. 

In Sections \ref{sec:map1} to \ref{sec:map3}, we consider only one symbol interval and do not consider ISI. We consider the ISI case in Section \ref{sec:isi}. 


\subsection{The MAP framework} 
\label{sec:map1} 
We adopt a MAP framework for detection. Let ${\mathbf P}[s | {\cal B}(t)]$ denote the posteriori probability that symbol $s$ has been sent given the history ${\cal B}(t)$. If the demodulation decision is to be done at time $t$, then the demodulator decides that symbol $\hat{s}$ has been sent if 
\begin{align}
\hat{s} = {\arg\max}_{s = 0, ..., K-1}  {\mathbf P}[s | {\cal B}(t)]
\end{align}

Instead of working with ${\mathbf P}[s | {\cal B}(t)]$, we will work with its logarithm. Let 
\begin{align}
L_s(t) = \log ({\mathbf P}[s | {\cal B}(t)])
\end{align}

The first step is to determine $L_s(t+\Delta t)$ from $L_s(t)$. Given ${\cal B}(t)$ is the section of $b(t)$ in the time interval $[0,t]$, one can consider ${\cal B}(t+\Delta t)$ as the concatenation of ${\cal B}(t)$ and the section of $b(t)$ in the time interval $(t,t + \Delta t]$. Over an infinitesimal $\Delta t$, we can consider the signal $b(t)$ is a constant in the time interval $(t,t + \Delta t]$; we therefore abuse the notation and use $b(t + \Delta t)$ to denote the section of $b(t)$ in the time interval $(t,t + \Delta t]$. By using Bayes' rule, it can be shown that 
\ifsingle
\begin{align}
L_s(t+\Delta t) =& L_s(t) + \log (\mathbf{P}[b(t+\Delta t) | s, {\cal B}(t)]) 
 - \log (\mathbf{P}[b(t+\Delta t) | {\cal B}(t)])
\label{eqn:logpp_prelim}
\end{align}
\else
\begin{align}
L_s(t+\Delta t) =& L_s(t) + \log (\mathbf{P}[b(t+\Delta t) | s, {\cal B}(t)]) \nonumber \\ 
& - \log (\mathbf{P}[b(t+\Delta t) | {\cal B}(t)])
\label{eqn:logpp_prelim}
\end{align}
\fi
where $\mathbf{P}[b(t+\Delta t) | s, {\cal B}(t)]$ is the probability that there are $b(t+\Delta t)$ complexes given that the transmitter has sent the symbol $s$ and the previous history ${\cal B}(t)$. The last term on the RHS of \eqref{eqn:logpp_prelim}, i.e. $\mathbf{P}[b(t+\Delta t) | {\cal B}(t)]$, is independent of the transmission symbol so we do not need it for the purpose of detection. We will focus on determining $\mathbf{P}[b(t+\Delta t) | s, {\cal B}(t)]$. 

%
%

\subsection{Computing $\mathbf{P}[b(t+\Delta t) | s, {\cal B}(t)]$}
The problem of determining the probability $\mathbf{P}[b(t+\Delta t) | s, {\cal B}(t)]$ is essentially a Bayesian filtering problem. Recall that the complete state of the system is $(N(t),b(t))$ and the receiver can only observe $b(t)$, therefore the task of the receiver is to use the history ${\cal B}(t)$ and the system model to do prediction. Standard method can be used to derive $\mathbf{P}[b(t+\Delta t) | s, {\cal B}(t)]$ but the derivation is long, especially because of the diffusion terms; the derivation can be found in Appendix \ref{app:proofA}. The result is  
\ifsingle
\begin{align}
& \mathbf{P}[b(t+\Delta t) | s, {\cal B}(t)] \nonumber \\ 
	= & \delta_{b(t+\Delta t), b(t) + 1} \; \lambda (M - b(t))  \; \Delta t  \;  \mathbf{E} [n_R(t) | s,  {\cal B}(t)]  +  
	    \delta_{b(t+\Delta t), b(t) - 1} \; \mu b(t) \; \Delta t \; + \nonumber  \\
	   & \delta_{b(t+\Delta t), b(t)} \;  
	     (1 - \lambda (M - b(t)) {\mathbf E}[n_R(t) | s,  {\cal B}(t)] \; \Delta t - \mu b(t)  \; \Delta t) 
	   \label{eqn:predictb}
\end{align}
\else
\begin{align} 
& \mathbf{P}[b(t+\Delta t) | s, {\cal B}(t)] \nonumber \\ 
	= & \delta_{b(t+\Delta t), b(t) + 1} \lambda (M - b(t))  \; \Delta t  \;  \mathbf{E} [n_R(t) | s,  {\cal B}(t)]  +  \nonumber \\
	   & \delta_{b(t+\Delta t), b(t) - 1} \mu b(t) \; \Delta t \; + \nonumber  \\
	   & \delta_{b(t+\Delta t), b(t)} \times \nonumber \\ 
	   &  (1 - \lambda (M - b(t)) {\mathbf E}[n_R(t) | s,  {\cal B}(t)] \; \Delta t - \mu b(t)  \; \Delta t) 
	   \label{eqn:predictb}
\end{align}
\fi
Note that only one of the three terms on the RHS of Equation \eqref{eqn:predictb} is non-zero depending on whether the observed $b(t+\Delta t)$ is one more, one less or equal to that of $b(t)$.
The term $\mathbf{E} [n_R(t) | s,  {\cal B}(t)]$ is the expected number of signalling molecules in the receiver voxel given the history and the symbol $s$. The meaning of this term is that the receiver uses the history to predict what the expected number of signalling molecules in the receiver voxel is. Note that only the chemical kinetic parameters $\lambda$ and $\mu$ of the receptor appear explicitly in Equation \eqref{eqn:predictb}. Other parameters, such as the set of chemical reaction that generate Symbol $s$ and the diffusion coefficient, do not appear explicitly in Equation \eqref{eqn:predictb} but influence the system behaviour via the term $\mathbf{E} [n_R(t) | s,  {\cal B}(t)]$. 

\subsection{The demodulation filter} 
\label{sec:map3} 
\begin{figure}
\begin{center}
\includegraphics[page=3,trim=0cm 0cm 0cm 0cm ,clip=true, width=\picwidth]{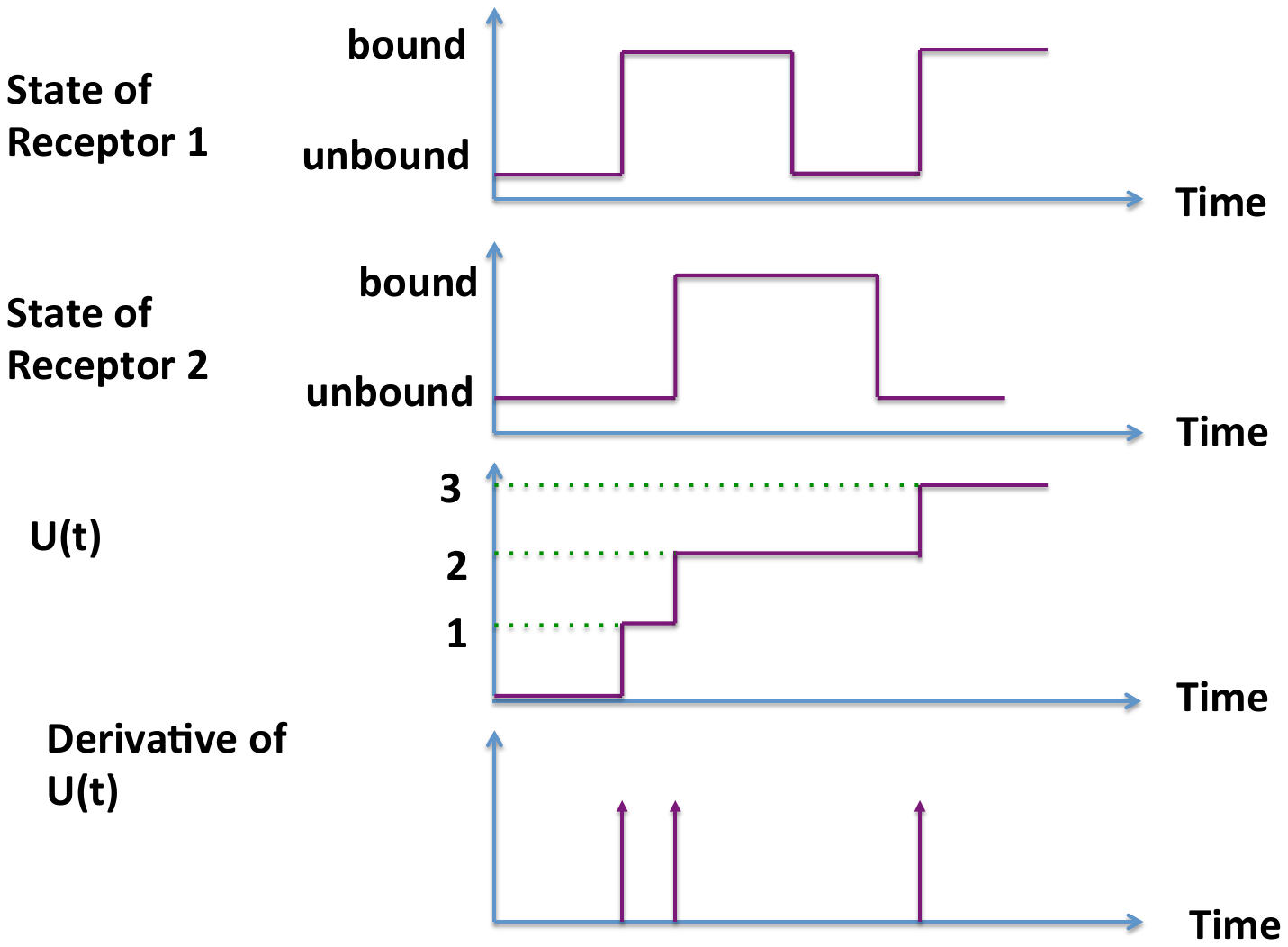}
\caption{This figure explains the meaning of the function $U(t)$, which is the total number of unbound-to-bound transitions for all receptors. }
\label{fig:ut}
\end{center}
\end{figure}

By substituting Equation \eqref{eqn:predictb} into Equation \eqref{eqn:logpp_prelim} and let $\Delta t$ go to zero, we show in Appendix \ref{app:proofB} that 
\ifsingle
\begin{align}
\frac{dL_s(t)}{dt} =& \frac{dU(t)}{dt} \log( {\mathbf E}[n_R(t) | s, {\cal B}(t)] ) - 
 \lambda (M - b(t)) {\mathbf E}[n_R(t) | s, {\cal B}(t)] + \tilde{L}(t) 
\label{eqn:logpp_dd} 
\end{align}
\else
\begin{align}
\frac{dL_s(t)}{dt} =& \frac{dU(t)}{dt} \log( {\mathbf E}[n_R(t) | s, {\cal B}(t)] ) - \nonumber \\ 
& \lambda (M - b(t)) {\mathbf E}[n_R(t) | s, {\cal B}(t)] + \tilde{L}(t) 
\label{eqn:logpp_dd} 
\end{align}
\fi
with $L_s(0)$ initialised to the logarithm of the prior probability that Symbol $s$ is sent. 
Equation \eqref{eqn:logpp_dd} is the {\sl optimal} demodulation filter. 
The term $U(t)$ is the cumulative number of times that the receptors have turned from the unbound to bound state. The meaning of $U(t)$ is illustrated in Figure \ref{fig:ut} assuming there are two receptors. The top two pictures in Figure \ref{fig:ut} show the state transitions for the two receptors. The third picture shows the function $U(t)$ which is increased by one every time a receptor switches from the unbound to bound state. The bottom picture shows $\frac{dU(t)}{dt}$ which is the derivative of $U(t)$. Note that $\frac{dU(t)}{dt}$ consists of a train of impulses (or Dirac deltas) where the timings of the impulses are the times at which a receptor binding event occurs. Loosely speaking, one may also view $\frac{dU(t)}{dt}$ as $\max(\frac{db(t)}{dt},0)$. 

The function $\tilde{L}(t)$, which is the last term on the RHS of \eqref{eqn:logpp_dd}, contains all the terms that are independent of Symbol $s$. Since $L_s(t)$ does not appear on the RHS of \eqref{eqn:logpp_dd}, this means that $\tilde{L}(t)$ adds the same contribution to all $L_s(t)$ for all $s = 0, ..., K-1$. We can therefore ignore $\tilde{L}(t)$ for the purpose of demodulation. 



The term ${\mathbf E}[n_R(t) | s, {\cal B}(t)]$ in Equation \eqref{eqn:logpp_dd} is the prediction of the mean number of signalling molecules in the receiver voxel using the history of receptor state. This is a filtering problem which requires extensive computation. Instead, we assume that the receiver has prior knowledge that if Symbol $s$ is transmitted, then the mean number of signalling molecules in the receiver voxel is $\sigma_s(t) = {\mathbf E}[n_R(t) | s]$ and the receiver uses $\sigma_s(t)$ for demodulation. We can view $\sigma_s(t)$ as internal models that the demodulator uses. The use of internal models is fairly common in signal processing and communication, e.g. a matched filter correlates the measured data with an expected response. 

\begin{figure}
\begin{center}
\includegraphics[page=5,trim=2cm 4cm 4cm 3cm ,clip=true, width=\picwidth]{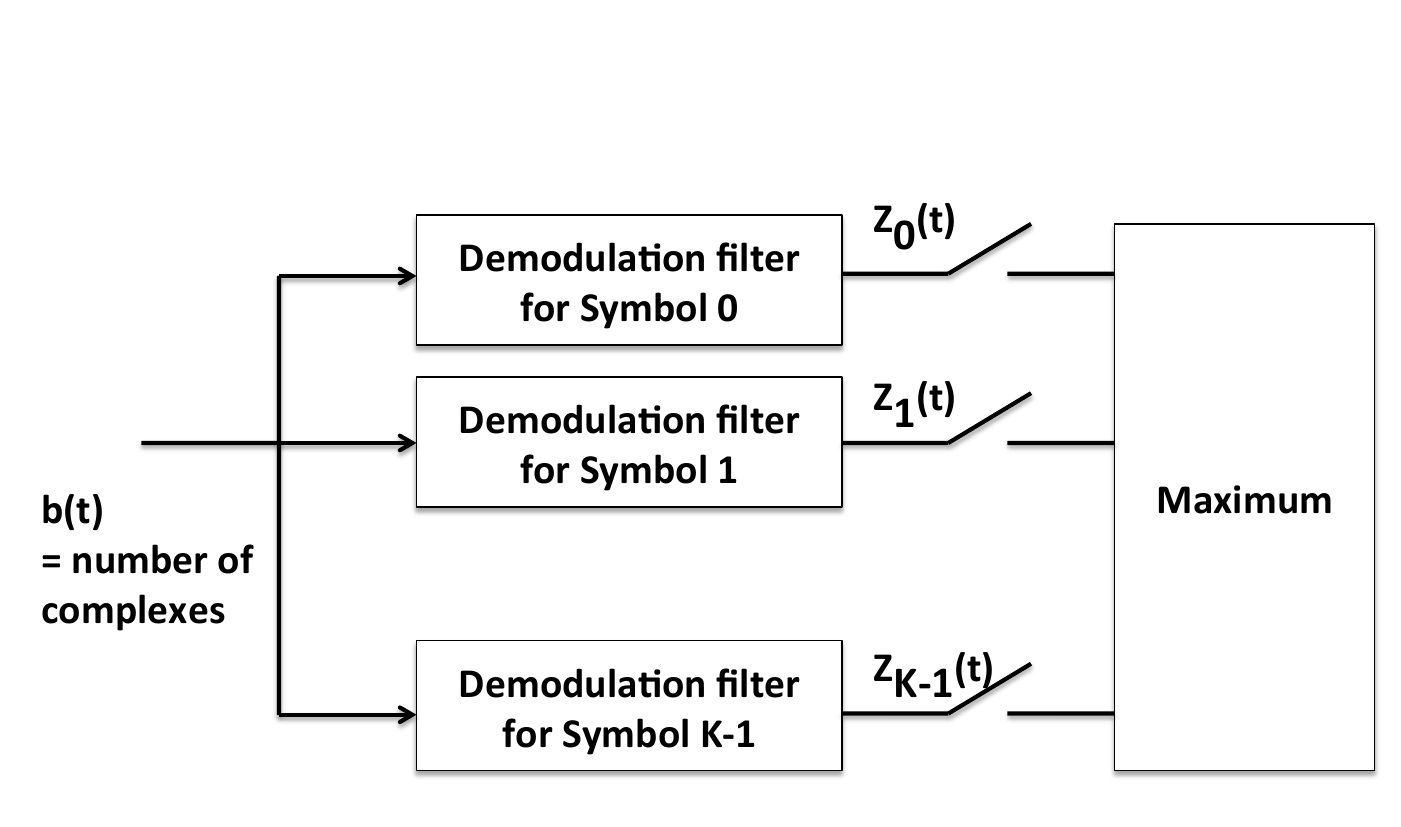}
\caption{The demodulator structure.}
\label{fig:demod}
\end{center}
\end{figure}

After making the modifications described in the last two paragraphs, we are now ready to describe the demodulator. Using $b(t)$ as the input, the demodulator runs the following $K$ continuous-time filters in parallel: 
\begin{align}
\frac{dZ_s(t)}{dt} = \frac{dU(t)}{dt} \log(\sigma_s(t)) - \lambda (M - b(t)) \sigma_s(t)
\label{eqn:logmap_s}
\end{align}
where $Z_s(0)$ is initialised to the logarithm of the prior probability that the transmitter sends Symbol $s$. If the demodulator makes the decision at time $t$, then the demodulator decides that Symbol $\hat{s}$ has been transmitted if 
\begin{align}
\hat{s} = {\arg\max}_{s = 0, ..., K-1} Z_s(t) 
\end{align}
The demodulator structure is illustrated in Figure \ref{fig:demod}. By comparing Equations \eqref{eqn:logpp_dd} and \eqref{eqn:logmap_s}, it can be shown that $L_{s_1}(t) - L_{s_2}(t) = Z_{s_1}(t) - Z_{s_2}(t)$ for any two symbols $s_1$ and $s_2$. An interpretation of the demodulation filter output $Z_s(t)$ is that $\exp(Z_s(t))$ is proportional to the posteriori probability ${\mathbf P}[s | {\cal B}(t)]$. 

Note that the replacement of ${\mathbf E}[n_R(t) | s, {\cal B}(t)]$ in Equation \eqref{eqn:logpp_dd} by $\sigma_s(t)$ means that the demodulation filter \eqref{eqn:logmap_s} is {\sl sub-optimal}. If ${\mathbf E}[n_R(t) | s, {\cal B}(t)]$ and $\sigma_s(t)$ are close to each other, we expect the performance degradation to be small. It is an open problem what the difference between ${\mathbf E}[n_R(t) | s, {\cal B}(t)]$ and $\sigma_s(t)$ is. The difficulty in answering this problem is due to the fact that ligand-receptor binding is a nonlinear process. In Appendix \ref{app:C}, we motivate the closeness between 
${\mathbf E}[n_R(t) | s, {\cal B}(t)]$ and $\sigma_s(t)$ by studying an analogous problem in linear time-invariant (LTI) systems. We will compare the performance of the optimal and sub-optimal demodulation filters in Section \ref{sec:prop_cf_opt_bf}. 



In order to understand Equation \eqref{eqn:logmap_s}, we consider the situation where Symbol 1 generates a lot more signalling molecules than Symbol 0 such that it results in more signalling molecules in the receiver voxel, or $\sigma_1(t) > \sigma_0(t)$ for all $t$. If the transmitter sends Symbol 1, then more signalling molecules are expected to reach the receiver voxel. The consequence is that there are more receptor binding events  and the number of unbound receptors $(M-b(t))$ is smaller. Therefore, in Equation \eqref{eqn:logmap_s}, we expect a big positive contribution from the first term on the RHS and a small negative contribution from the second term. The net effect is a big $Z_1(t)$. On the other hand, if the transmitter sends Symbol 0, the number of receptor binding events is smaller and $(M-b(t))$ is big. This results in a smaller $Z_0(t)$. Therefore, $Z_1(t)$ is likely to be bigger than $Z_0(t)$, which means correct detection.  

\subsection{Discussions} 
\subsubsection{Implementation issues} 
 The implementation of the demodulator is an open research problem. The demodulation filter \eqref{eqn:logmap_s} is an analogue filter and it requires the internal model $\sigma_i(t)$. The design of analogue circuits using chemical reactions for calculations is an active research area, see \cite{Sarpeshkar:2014dg} for a recent overview. The demodulation filter requires logarithm, multiplication, subtraction, integration and counting the number of times the receptors have switched from the unbound to bound state. An analogue molecular circuit for calculating logarithm is presented in \cite{Daniel:2013ke}. There are also circuits that can perform multiplication and subtraction \cite{Sarpeshkar:2014dg}. Living cells are known to use integration \cite{Yi:2000tj}. It may be possible to implement the counting using a chemical reaction \cite{Siggia:2013dd}. It may be possible to approximate the internal models by using some lower order chemical reactions. This discussion shows that some components to implement the demodulator exist but the exact implementation remains an open problem. 

In order to bypass the difficulty of computing ${\mathbf E}[n_R(t) | s, {\cal B}(t)]$, we have proposed to use internal models $\sigma_s(t)$. An open research problem is to study sub-optimal estimation of ${\mathbf E}[n_R(t) | s, {\cal B}(t)]$. 

Note that \cite{Sarpeshkar:2014dg} shows that analogue computation is more efficient than digital computation if high precision is not required. An interesting problem is to study the impact of low precision analogue calculations on the demodulation performance. 

\subsubsection{Continuous transmission medium versus voxels}
\label{sec:map:css} 
We mention in Section \ref{sec:related} that some papers in molecular communications assume a continuous medium (rather than discretising the medium into voxels) and a non-zero receiver size. If a continuous medium is assumed, the state of a signalling molecule in the transmission medium is its position. Let $p(\vec{x},t)$ be the probability that a molecule is at position $\vec{x}$ at time $t$. The time evolution of $p(\vec{x},t)$ can again be modelled by a CTMP, which is continuous in both time and space. The time evolution of $p(\vec{x},t)$ is described by the differential Chapman-Kolmogorov Equation (CKE) \cite{Gardiner}. It is possible to derive an alternative CTMP by replacing the diffusion of signalling molecules in Equation \eqref{eqn:tp:eta} by differential CKE as well as by adding equations to describe how the signalling molecules enter or leave the receiver voxel. We can then apply Bayesian filtering to this alternative CTMP. We expect this alternative CTMP will give the same demodulator filter because in our derivation based on discrete voxels, the MAP demodulator depends only on the number of the signalling molecules in the receiver voxel and the diffusion parameters do not appear explicitly in the demodulation filter. Further support of this argument is given in the derivation in Appendix \ref{app:proofB} where we show that the $d_{ij}$ parameters in \eqref{eqn:tp:eta} are cancelled out in deriving the Bayesian filter.  


\subsection{Inter-symbol interference}
\label{sec:isi} 

It is in principle possible to use the demodulation filters \eqref{eqn:logmap_s} to deal with the case with ISI. We use $T_x$ to denote one symbol duration and we assume that ISI only lasts for a finite amount of time. Specifically, consider a symbol sent in $[kT_x,(k+1)T_x]$, we assume that the effect of this transmission can be neglected after the time $ (k+n) T_x$. This can be realised by appropriately choosing the transmitter and receiver parameters, $T_x$ and $n$.

In order to make the explanation here a bit more concrete, we assume that the transmitter uses $K = 2$ symbols and $n = 3$. Over a duration of $n = 3$ symbols, the possible sequences sent by the transmitter are 000, 001, 010, 011, ... , 111. Let $\sigma_{0,0,0}(t)$ denote the mean number of signalling molecules at the receiver voxel if the sequence 000 is sent. We can similarly define $\sigma_{0,0,1}(t), ..., \sigma_{1,1,1}(t)$. Consider the transmission of three consecutive symbols $s_{k-2}, s_{k-1}$ and $s_k$. Assuming that we have an estimation of the first two symbols $\hat{s}_{k-2}$ and $\hat{s}_{k-1}$, then the decoding of $s_k$ can be done by using the demodulation filter \eqref{eqn:logmap_s} by replacing $\sigma_s(t)$ by $\sigma_{\hat{s}_{k-2},\hat{s}_{k-1},s}$. For example, if  $\hat{s}_{k-2} = 1$ and $\hat{s}_{k-1} = 0$, then one can decode what $s_k$ is by using the demodulator filters $\sigma_{1,0,1}(t)$ and $\sigma_{1,0,1}(t)$. Although the decision feedback based method can solve the ISI problem, the number of internal models increases exponentially with the memory length parameter $n$. 

The reason why we need to consider all $2^n$ possible transmission sequences is that the ligand-receptor binding process has a non-linear reaction rate. A method to reduce the number of internal models is to design the system so that $\sigma_{0,0,0}(t)$ etc. can be decomposed into a sum. Let $\sigma_s(t)$ ($s = 0,1$) be the mean number of signalling molecules at the receiver voxel if the symbol $s$ is sent for one symbol duration and in the absence of ISI. If 
\begin{align}
\sigma_{s_1,s_2,s_3}(t) \approx \sigma_{s_1}(t - 2 T_x) + \sigma_{s_2}(t - T_x) + \sigma_{s_3}(t) 
\label{eqn:sigma_sum} 
\end{align}
holds for all $s_1, s_2$ and $s_3$, then one can again make use of decision feedback to decode the ISI signal. However, this time, only $K$ internal models are needed. Equation \eqref{eqn:sigma_sum} can be made to hold approximately if the number of receptors is large. This can be explained as follows. First of all, if ligand-receptor binding is absent, this means there is only free diffusion then Equations \eqref{eqn:sigma_sum} holds because the mean number of signalling molecules obeys the diffusion equation which is linear. This means that we need to create an environment that ``looks like" free diffusion even when ligand-receptor binding is present. This can be realised if the number of signalling molecules that are bound to the receptors is small compared to those that are free. A method to achieve this is to increase the number of receptors. We will demonstrate this with a numerical example in Section \ref{sec:eval}. However, it is still an open problem to solve the ISI in the general case.

\section{Properties of the demodulator}
\label{sec:eval}
The aim of this section is to study the properties of the MAP demodulator numerically. We begin with the methodology.

\subsection{Methodology} 
We assume the diffusion coefficient $D$ of the medium is 1 $\mu$m$^2$s$^{-1}$. The receptor parameters are $\tilde{\lambda}$ = 0.005 $\mu$m$^3$ s$^{-1}$, $\lambda = \frac{\tilde{\lambda}}{W^3}$, and $\mu = 1$ s$^{-1}$. These values are similar to those used in \cite{Erban:2009us} and 
\cite{Pierobon:2011ve}~\footnote{
The paper \cite{Pierobon:2011ve} considers ligand-receptor binding in the chemical master equation setting. In our notation, the parameter values in \cite{Pierobon:2011ve} are $D$ = 100 $\mu$m$^2$s$^{-1}$, $\tilde{\lambda}$ = 0.2 $\mu$m$^3$ s$^{-1}$ and $\mu = 10$ s$^{-1}$. These parameters are 10--100 times faster than ours and can be considered as a time-scaling. Note that \cite{Pierobon:2011ve} uses $k_+$ and $k_-$ instead of, respectively, $\tilde{\lambda}$ and $\mu$. 
}. 
The above parameter values will be used for all the numerical experiments. 

For each experiment, the transmitter uses either $K = 2$ or $K = 3$ symbols. Each symbol is generated by a different sets of chemical reactions. Different experiments may use different sets of chemical reactions and will be described later. The number of receptors also varies between the experiments. 

We use the Stochastic Simulation Algorithm (SSA) \cite{Gillespie:1977ww} to obtain realisations of $b(t)$ which is the number of complexes over time. SSA is a standard algorithm in chemistry to simulate diffusion and reactions; it is essentially an algorithm to simulate a CTMP. 

In order to use Equation \eqref{eqn:logmap_s}, we require the mean number of signalling molecules $\sigma_s(t)$ in the receiver voxel when Symbol $s$ is sent. Unfortunately, it is not possible to analytically compute $\sigma_s(t)$ from the CTMP because of moment closure problem which arises when the transition rate is a non-linear function of the state \cite{Smadbeck:2013fk}. We therefore resort to simulation to estimate $\sigma_s(t)$. Each time when we need an $\sigma_s(t)$, we run SSA simulation 500 times and average the results to obtain $\sigma_s(t)$. Note that these simulations are different from those that we use to generate $b(t)$ for the performance study. In other words, the simulations for estimating $\sigma_s(t)$ and for performance study are completely independent. 

Once $b(t)$ and $\sigma_s(t)$ are obtained, we use numerical integration to calculate $Z_s(t)$ using Equation \eqref{eqn:logmap_s}. We assume that all symbols appear with equal probability, so we initialise $Z_s(0) = 0$ for all $s$. 

\subsection{Optimal filter \eqref{eqn:logpp_dd} versus sub-optimal filter \eqref{eqn:logmap_s}}
\label{sec:prop_cf_opt_bf}
The optimal demodulation filter \eqref{eqn:logpp_dd} requires the term ${\mathbf E}[n_R(t) | s, {\cal B}(t)]$ which can be obtained by solving a Bayesian  filtering problem. Since filtering problems are computationally expensive to solve, we propose the sub-optimal demodulation filter \eqref{eqn:logmap_s} which uses $\sigma_s(t) = {\mathbf E}[n_R(t) | s]$ as an internal model. The aim of this section is to compare the performance of these two demodulation filters. 

In this comparison, we consider a medium of 1$\mu$m $\times$ $\frac{1}{3}$$\mu$m $\times$ $\frac{1}{3}$$\mu$m. We assume a voxel size of ($\frac{1}{3}$$\mu$m)$^{3}$ (i.e. $W = \frac{1}{3}$ $\mu$m), creating an array of $3 \times 1 \times 1$ voxels. The voxel co-ordinates of transmitter and receiver are, respectively, (1,1,1) and (3,1,1). A reflecting boundary condition is assumed. 

The reason why we have chosen to use such a small number of voxels is because of the dimensionality of the filtering problem. For example, if each voxel can have a maximum of 100 signalling molecules at a time, then there are 10$^6$ possible $N(t)$ vectors and the filtering problem has to estimate the probability ${\mathbf P}[N(t) | s, {\cal B}(t)]$ for each possible $N(t)$ vector. Although there are approximation techniques to solve the Bayesian filtering problem, that would introduce inaccuracies. The use of small number of voxels will allow us to compute ${\mathbf P}[n_R(t) | s, {\cal B}(t)]$ precisely. 

For this experiment, we use $K = 2$ symbols and two values of $M$ (the number of receptors): 5 and 10. Both Symbols 0 and 1 use Reaction \eqref{cr:rna} such that Symbols 0 and 1 causes, respectively, 10 and 50 signalling molecules to be generated per second on average by the transmitter. The simulation time is about 1.8 seconds. 

\begin{figure}
\begin{center}
\includegraphics[width=\picwidth]{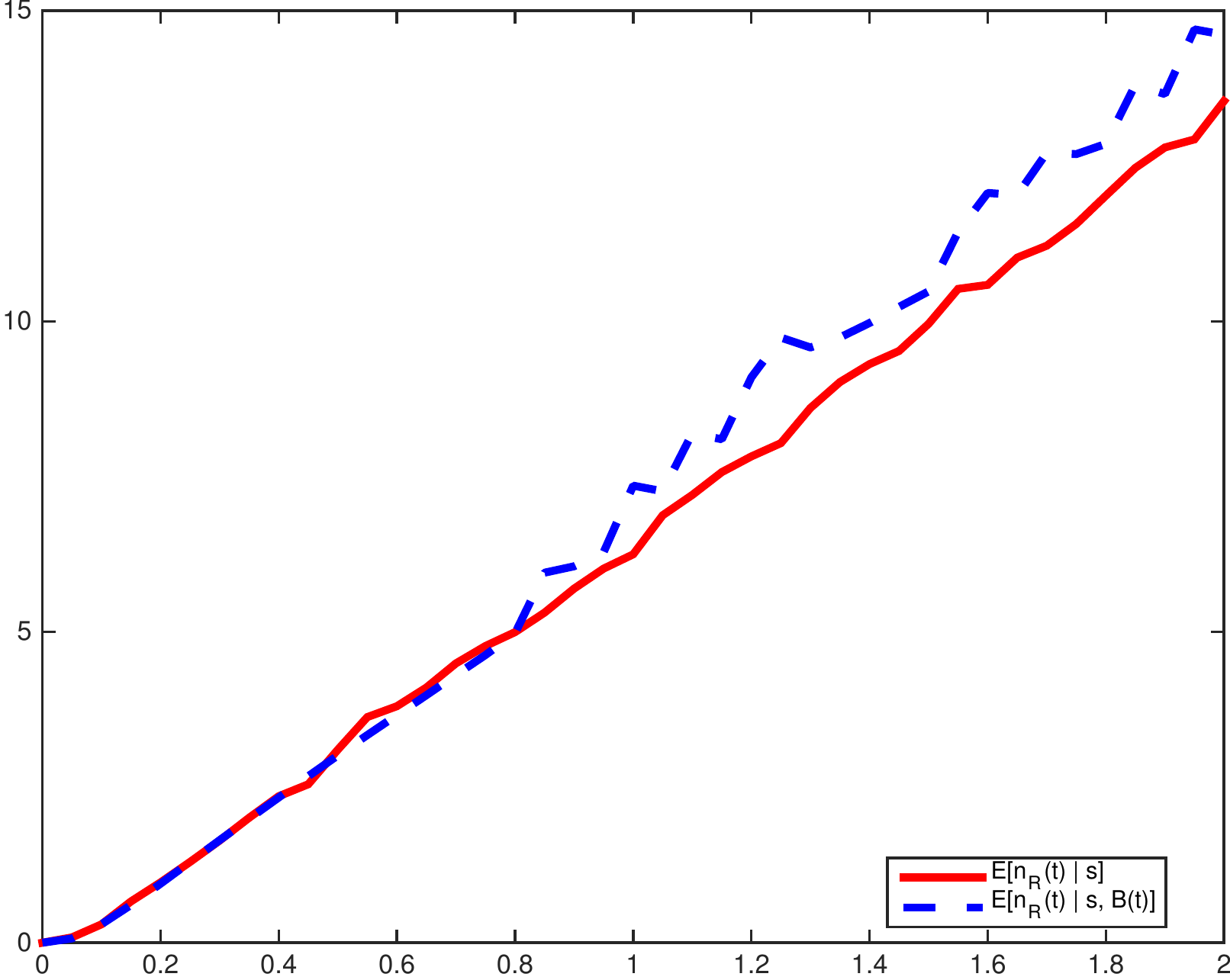}
\caption{The figure compares $\sigma_s(t)$ =  ${\mathbf E}[n_R(t) | s]$  and  ${\mathbf E}[n_R(t) | s, {\cal B}(t)]$ for one realisation of ${\cal B}(t)$.}
\label{fig:prop:opt_mean}
\end{center}
\end{figure}

We first show that ${\mathbf E}[n_R(t) | s, {\cal B}(t)]$ and $\sigma_s(t)$ are rather similar. Figure \ref{fig:prop:opt_mean} shows $\sigma_s(t)$ and a trajectory of ${\mathbf E}[n_R(t) | s, {\cal B}(t)]$ (obtained from one realisation of $b(t)$) are pretty similar. This result is obtained from using $M = 10$ and Symbol 1. The results for other choices of $M$, transmission symbols or other realisations of $b(t)$ are similar. 

\begin{figure}
\begin{center}
\includegraphics[width=\picwidth]{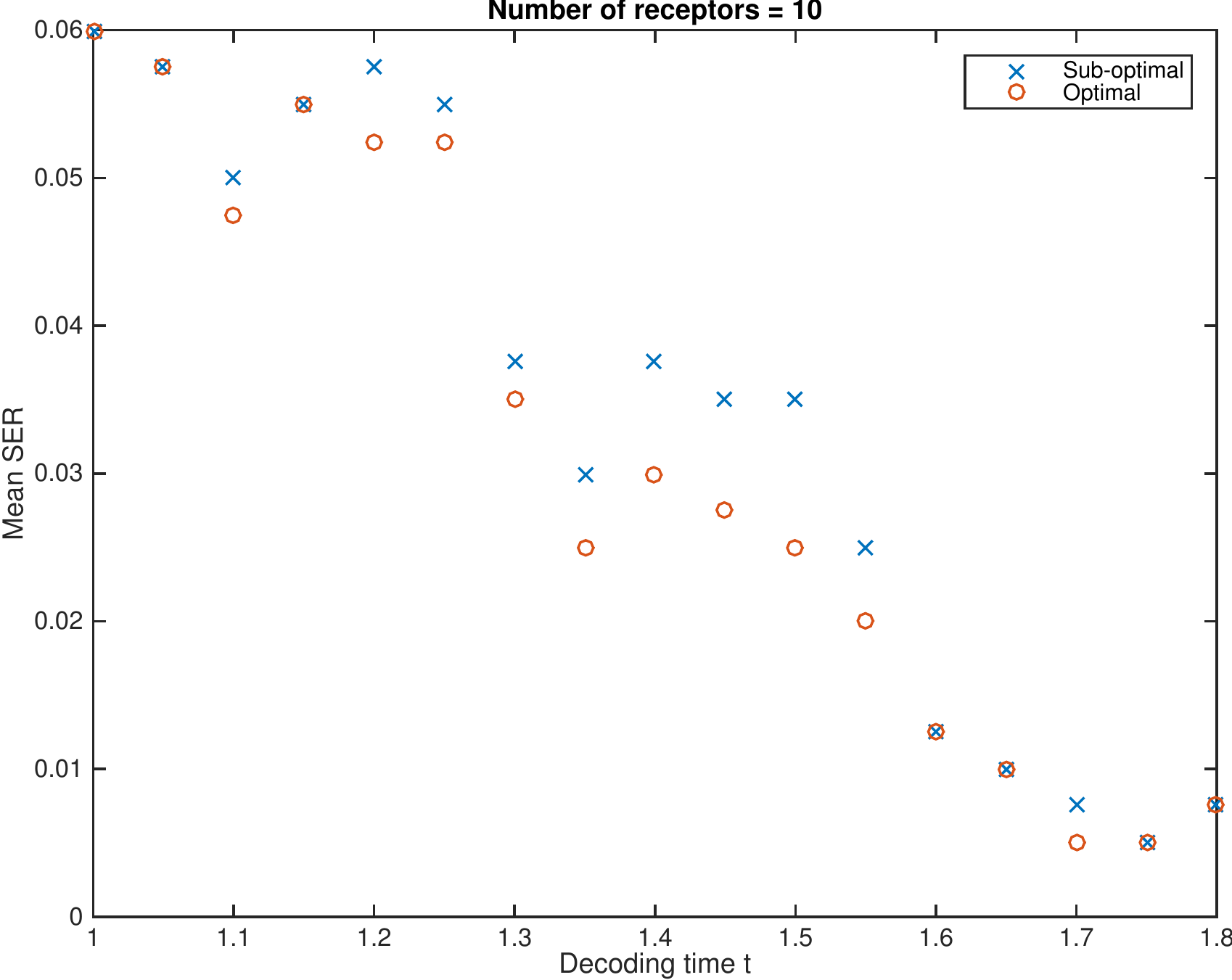}
\caption{The SERs of the optimal and sub-optimal filters for different decoding time $t$.}
\label{fig:prop:opt_ser}
\end{center}
\end{figure}

Figure \ref{fig:prop:opt_ser} shows the mean symbol error rates (SERs), for both optimal and sub-optimal demodulation filters, if the detection is done at time $t$ = 1, 1.05, 1.1, ..., 1.8. The SERs is obtained from 400 realisations of $b(t)$. The difference in SERs between the optimal and sub-optimal filter is less than 1\%. We have also checked that the two demodulators make the same decoding decision on average 99.3\% of the time. 

In the rest of this section, we will use the sub-optimal demodulation filter  \eqref{eqn:logmap_s} because of its lower computational complexity. 

\subsection{Properties of the demodulator output} 
\label{sec:prop_demod}
We consider a medium of 2$\mu$m $\times$ 2$\mu$m $\times$ 1 $\mu$m. We assume a voxel size of ($\frac{1}{3}$$\mu$m)$^{3}$ (i.e. $W = \frac{1}{3}$ $\mu$m), creating an array of $6 \times 6 \times 3$ voxels. The transmitter and receiver are located at (0.5,0.8,0.5) and (1.5,0.8,0.5) (in $\mu$m) in the medium. The voxel co-ordinates are (2,3,2) and (5,3,2) respectively. We assume an absorbing boundary for the medium and the signalling molecules escape from a boundary voxel surface at a rate of $\frac{d}{50}$. This configuration will be used for the rest of this section. 

For this experiment, we use $K = 2$ symbols and $M = 50$ receptors. Both Symbols 0 and 1 use Reaction \eqref{cr:rna} such that 
Symbols 0 and 1 causes, respectively, 40 and 80 signalling molecules to be generated per second on average by the transmitter. The simulation time is about 3 seconds. 

\begin{figure}
\begin{center}
\includegraphics[width=\picwidth]{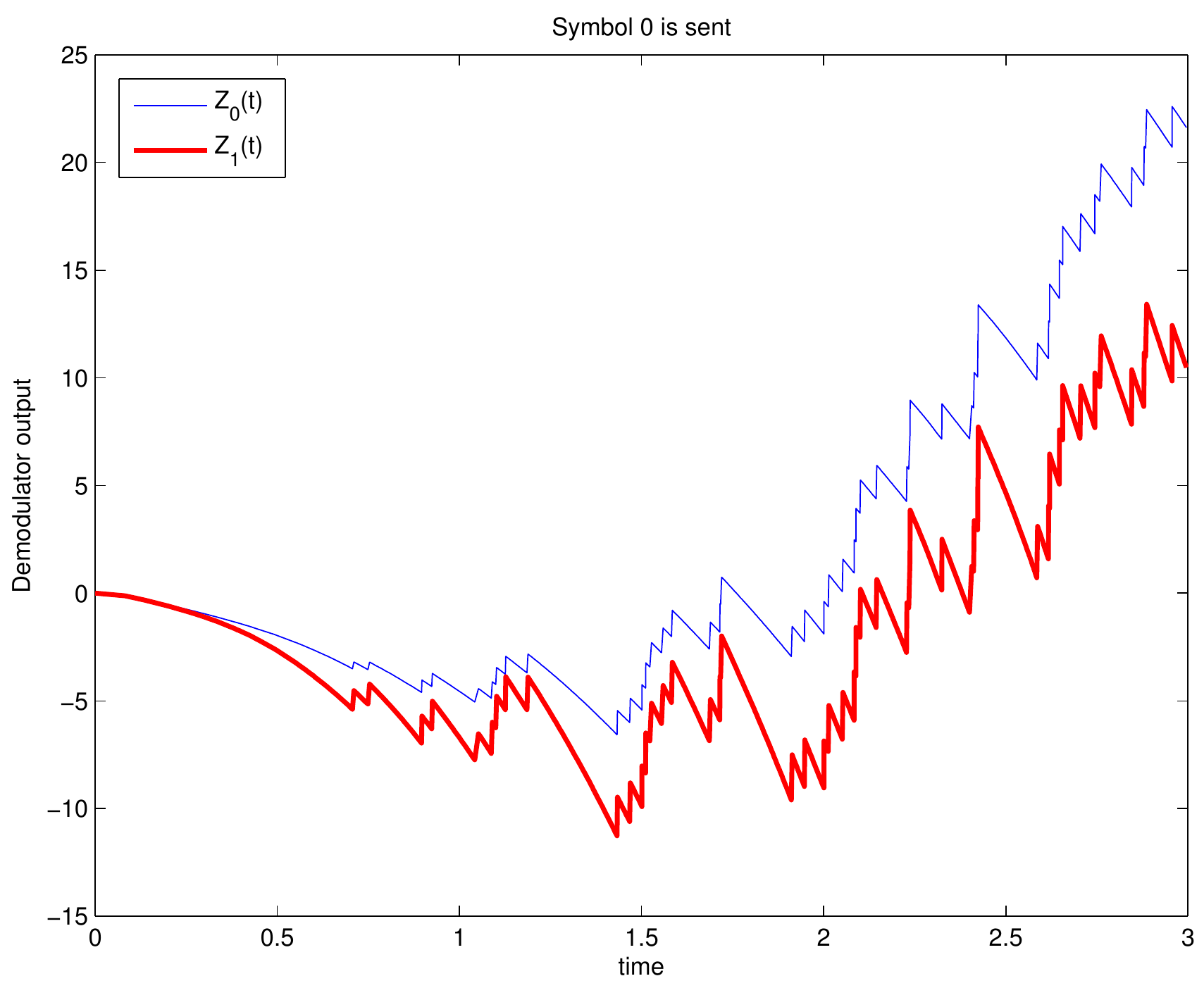}
\caption{The output of the modulators $Z_0(t)$ (thin line) and $Z_1(t)$ (thick line) for Symbol 0.}
\label{fig:prop_demod_raw_u0}
\end{center}
\end{figure}

\begin{figure}
\begin{center}
\includegraphics[width=\picwidth]{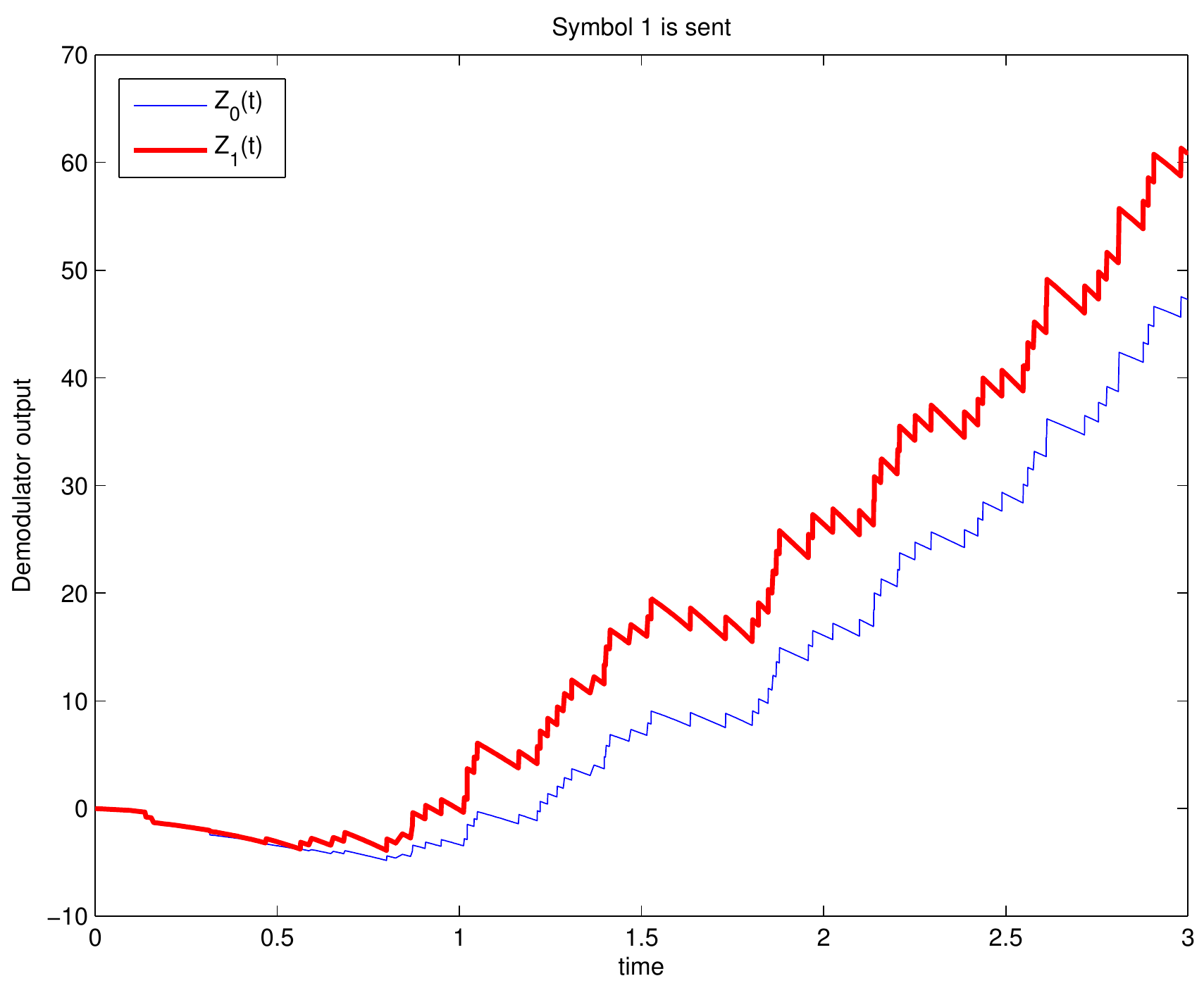}
\caption{The output of the modulators $Z_0(t)$ (thin line) and $Z_1(t)$ (thick line) for Symbol 1.}
\label{fig:prop_demod_raw_u1}
\end{center}
\end{figure}

Figure \ref{fig:prop_demod_raw_u0} shows the demodulation filter outputs $Z_0(t)$ and $Z_1(t)$ if the transmitter sends a Symbol 0. It can be seen that $Z_0(t) > Z_1(t)$ most of the time after $t = 1.2$, which means the detection is likely to be correct after this time. The sawtooth like appearance of $Z_0(t)$ and $Z_1(t)$ is due to the fact that every time when a receptor is bound, there is a jump in the filter output according to Equation \eqref{eqn:logmap_s}. Figure \ref{fig:prop_demod_raw_u1} shows the filter outputs $Z_0(t)$ and $Z_1(t)$ if the transmitter sends a Symbol 1; the behaviour is similar. 

\begin{figure}
\begin{center}
\includegraphics[width=\picwidth]{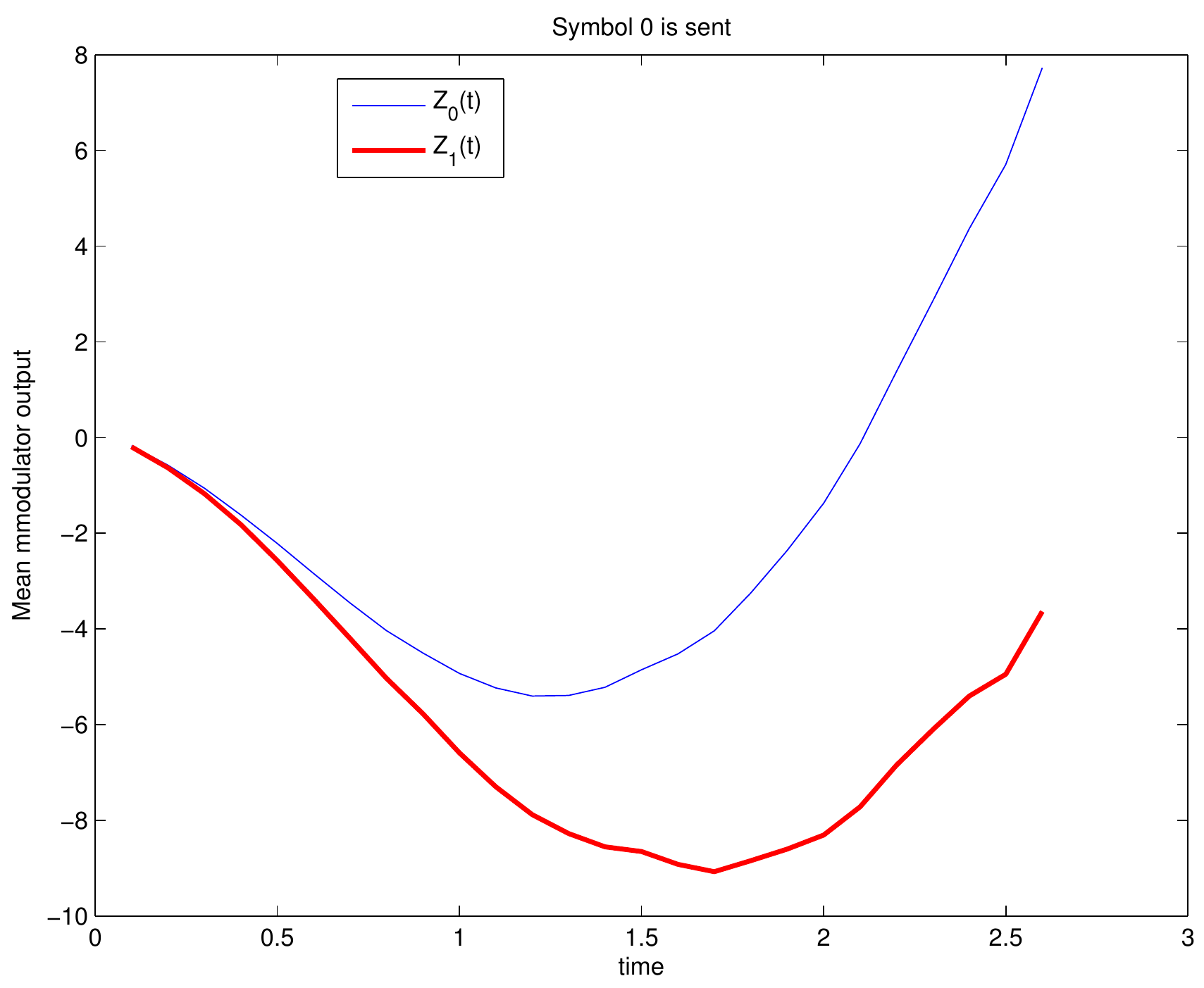}
\caption{The mean output of the modulators $Z_0(t)$ (thin line) and $Z_1(t)$ (thick line) for Symbol 0.}
\label{fig:prop_demod_mean_u0}
\end{center}
\end{figure}

Figure \ref{fig:prop_demod_mean_u0} shows the {\sl mean} filter outputs $Z_0(t)$ and $Z_1(t)$ if the transmitter sends a Symbol 0. The mean is computed over 200 realisations of $b(t)$. It can be seen that the mean filter output of $Z_0(t)$ is greater than that of $Z_1(t)$. Similarly, if Symbol 1 is sent, then we expect of the mean of $Z_1(t)$ to be bigger. The figure is not shown for brevity. 

\begin{figure}
\begin{center}
\includegraphics[width=\picwidth]{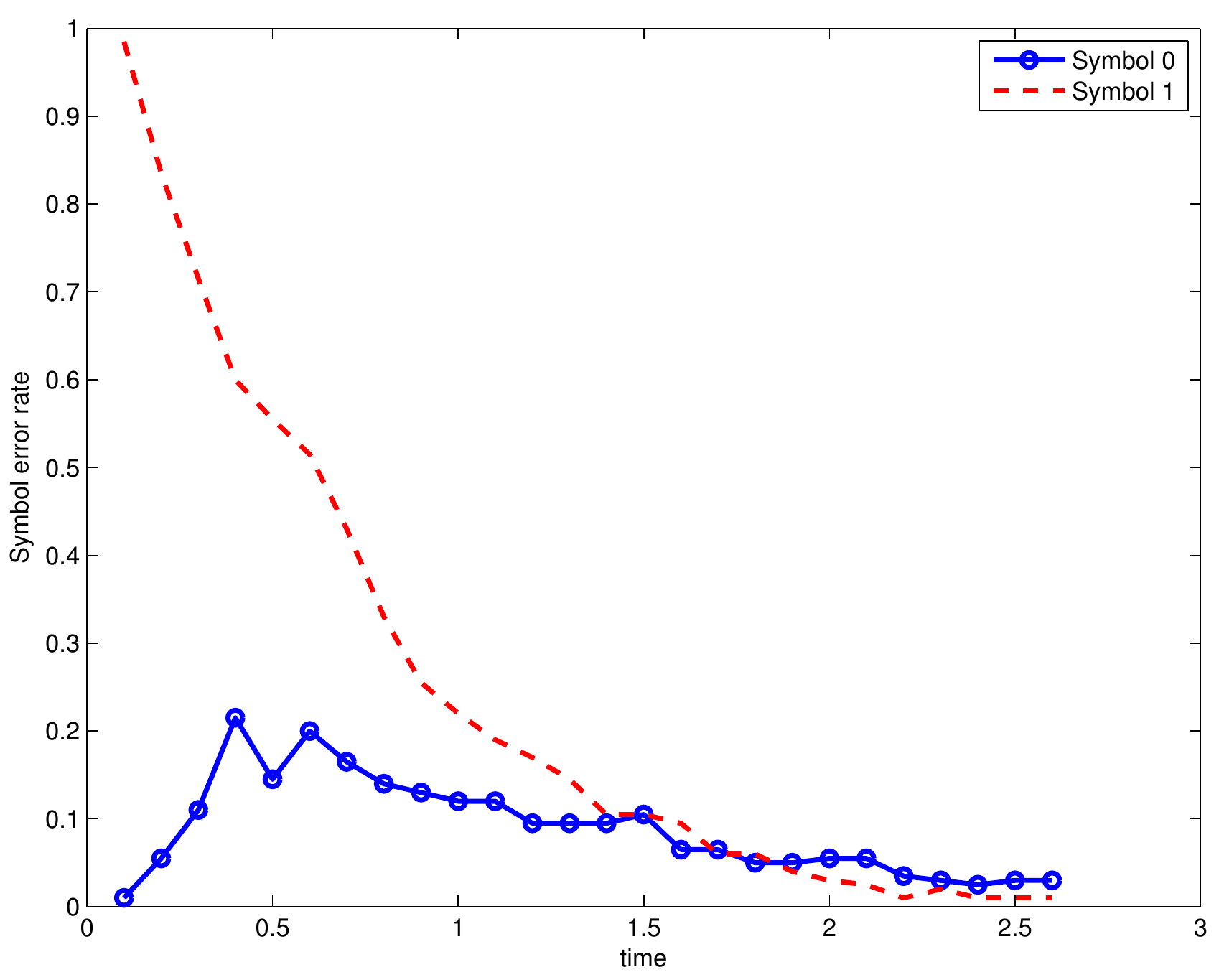}
\caption{The SER for Symbols 0 and 1.}
\label{fig:prop_demod_mean_ser}
\end{center}
\end{figure}

Figure \ref{fig:prop_demod_mean_ser} shows the mean SERs for Symbols 0 and 1 if the detection is done at time $t$. The SER for Symbol 1 is high initially but as more information is processed over time, the SER drops to a low value. 
This experiment shows that it is possible to use the analogue demodulation filter \eqref{eqn:logmap_s} to compute a quantity that allows us to distinguish between two emission patterns at the receiver.

%
%
%

\subsection{Impact of number of receptors}  
\begin{figure}
\begin{center}
\includegraphics[width=\picwidth]{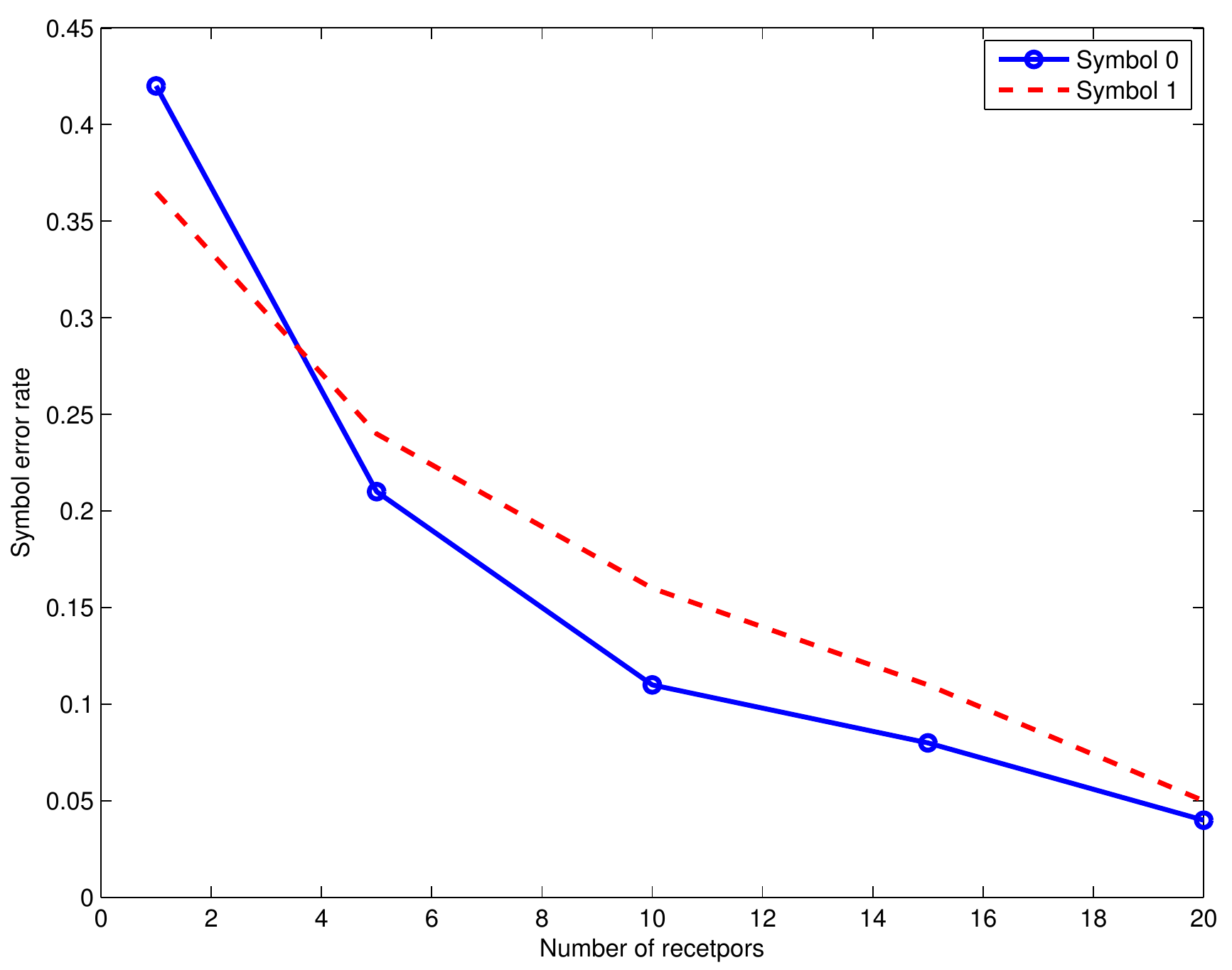}
\caption{The SER for Symbols 0 and 1 for varying number of receptors.}
\label{fig:prop_nrec}
\end{center}
\end{figure}

We continue with the setting of \ref{sec:prop_demod} but we vary the number of receptors between 1 and 20. We assume the demodulator makes the decision at $t = 2.5$ and calculate the mean SER for both symbols at $t = 2.5$. Figure \ref{fig:prop_nrec} plots the SERs versus the number of receptors. It can be seen that the SER drops with increasing number of receptors. 

\begin{figure}
\begin{center}
\includegraphics[width=\picwidth]{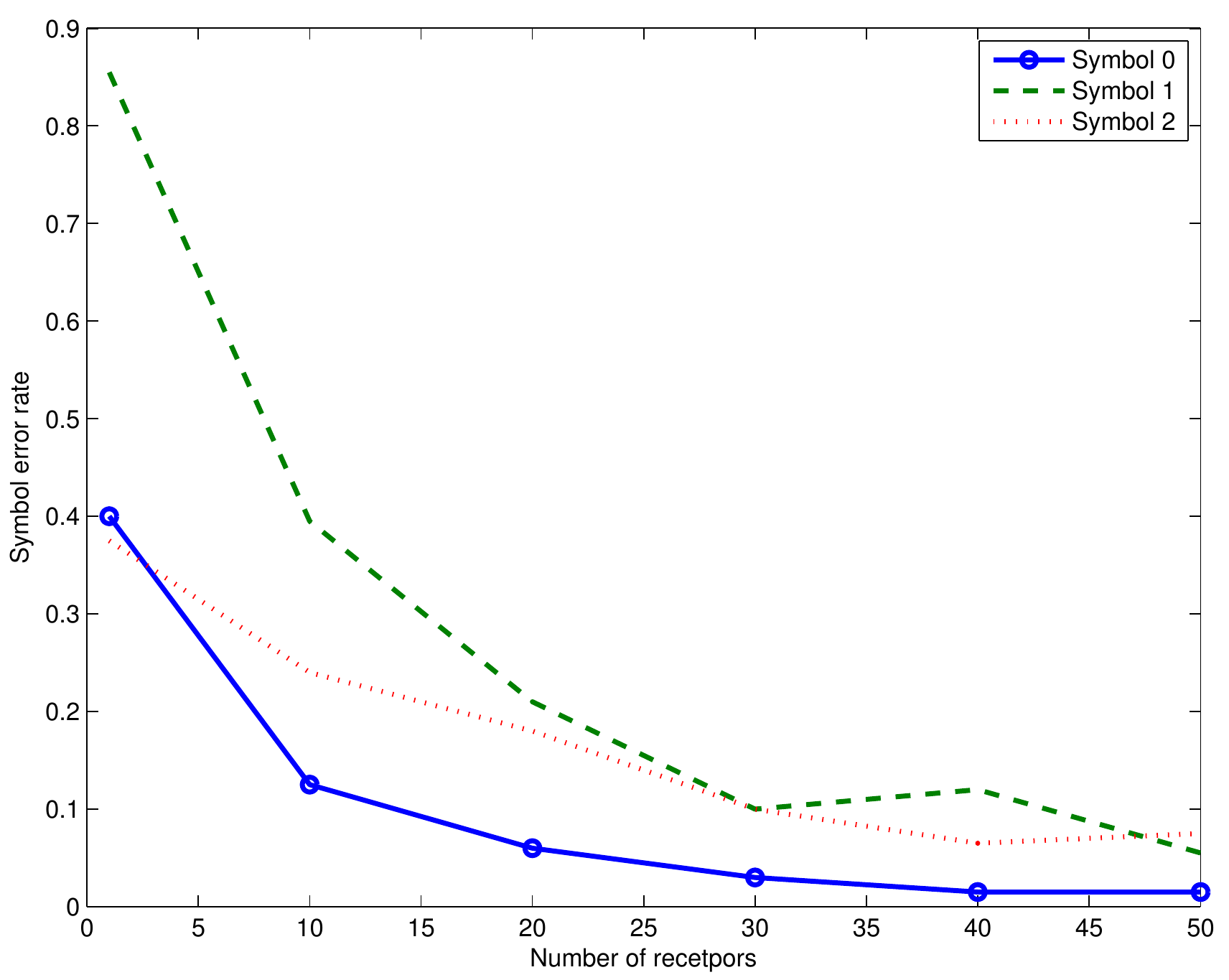}
\caption{The SER for Symbols 0, 1 and 2 for varying number of receptors.}
\label{fig:prop_nrec_3s}
\end{center}
\end{figure}

We have used $K = 2$ symbols so far. We retain the current Symbols 0 and 1, and add a Symbol 2 which is also of the form of Reaction \eqref{cr:rna} but its mean rate of production of signalling molecules is 3 times that of Symbol 0. The number of receptors $M$ used are: 1, 10, 20, ..., 150. 
 We compute the average SER at $t = 2.5$ assuming each symbol is transmitted with equal probability. We plot the logarithm of the average SER against $\log(M)$ in Figure \ref{fig:prop_nrec_3s}. 
 It can be seen that the SER drops with increasing number of receptors $M$. The plot in Figure \ref{fig:prop_nrec_3s} suggests that, when the number of receptors $M$ is large, the relationship between logarithm of SER and $\log(M)$ is linear. We perform a least-squares fit for $M$ between 50 and 150. The fitted straight line is shown in Figure \ref{fig:prop_nrec_3s} and it has a slope of $-1.13$. A possible explanation is that, because the receptors are non-interacting, each receptor provides an independent observation. The empirical evidence suggests that the average SER scales according to $\frac{1}{M}$ asymptotically provided that the voxel volume can contain that many receptors.

\subsection{Distinguishability of different chemical reactions}
\begin{figure}
\begin{center}
\includegraphics[width=\picwidth]{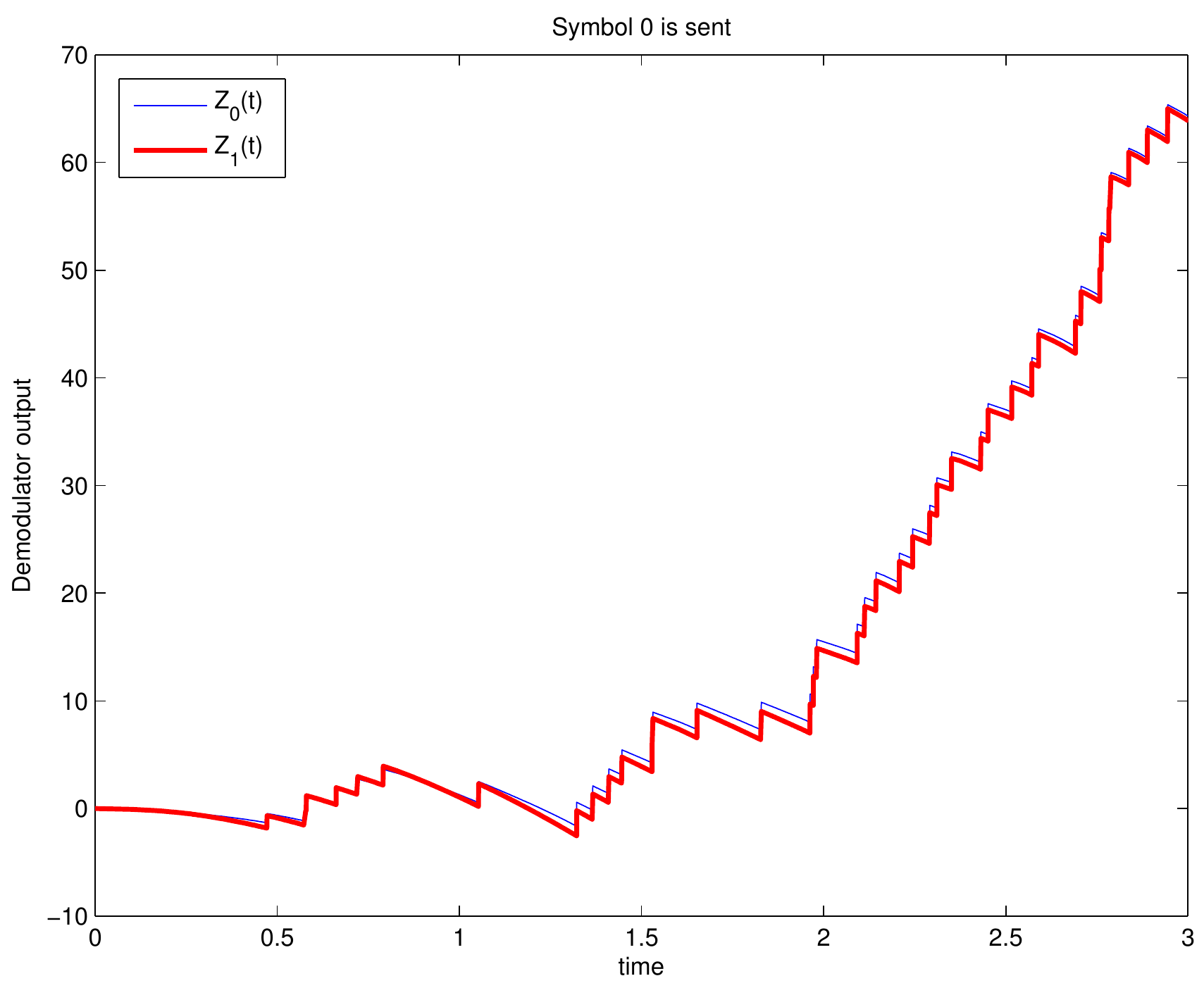}
\caption{The output of the modulators $Z_0(t)$ (thin line) and $Z_1(t)$ (thick line) for Symbol 0. The mean number of signalling molecules at the receiver voxel for both symbols is similar.}
\label{fig:equal}
\end{center}
\end{figure}

Equation \eqref{eqn:logmap_s} suggests that if the transmitter uses two sets of reactions which have almost the same mean number of signalling molecules in the receiver voxel, then it may be difficult to distinguish between these two symbols. In this study, Symbol 0 is generated by Reaction \eqref{cr:rna} with a rate of $\kappa$ while Symbol 1 is generated by:
\begin{align}
\cee{
[RNA]_{\rm ON} &  <->  [RNA]_{\rm OFF}
} \\
\cee{
[RNA]_{\rm ON}  &  ->[2\kappa] [RNA]_{\rm ON} + S 
} 
\label{eqn:r1} 
\end{align}
where we assume that RNA can be in an ON or OFF state, and signalling molecules $S$ are only produced when the RNA is in the ON-state. We assume that the there is an equal probability for the RNA to be in the two states and the reaction rate constant for the production of signalling molecule $S$ from ${\rm [RNA]}_{\rm ON}$ is $2\kappa$. This means that the mean rate of production of signalling molecules $S$ by Symbols 0 and 1 are the same. This gives rise to very similar $\sigma_0(t)$ and $\sigma_1(t)$. Figure \ref{fig:equal} shows the demodulation filter outputs $Z_0(t)$ and $Z_1(t)$ for one simulation. It can be seen that the two outputs are almost indistinguishable. Consequently, the SER is pretty high. This shows that symbols generated by reactions which have similar mean number of signalling molecules at the receiver voxel can be hard to distinguish. 

\subsection{Inter-symbol interference}
\begin{figure}
\begin{center}
\includegraphics[width=\picwidth]{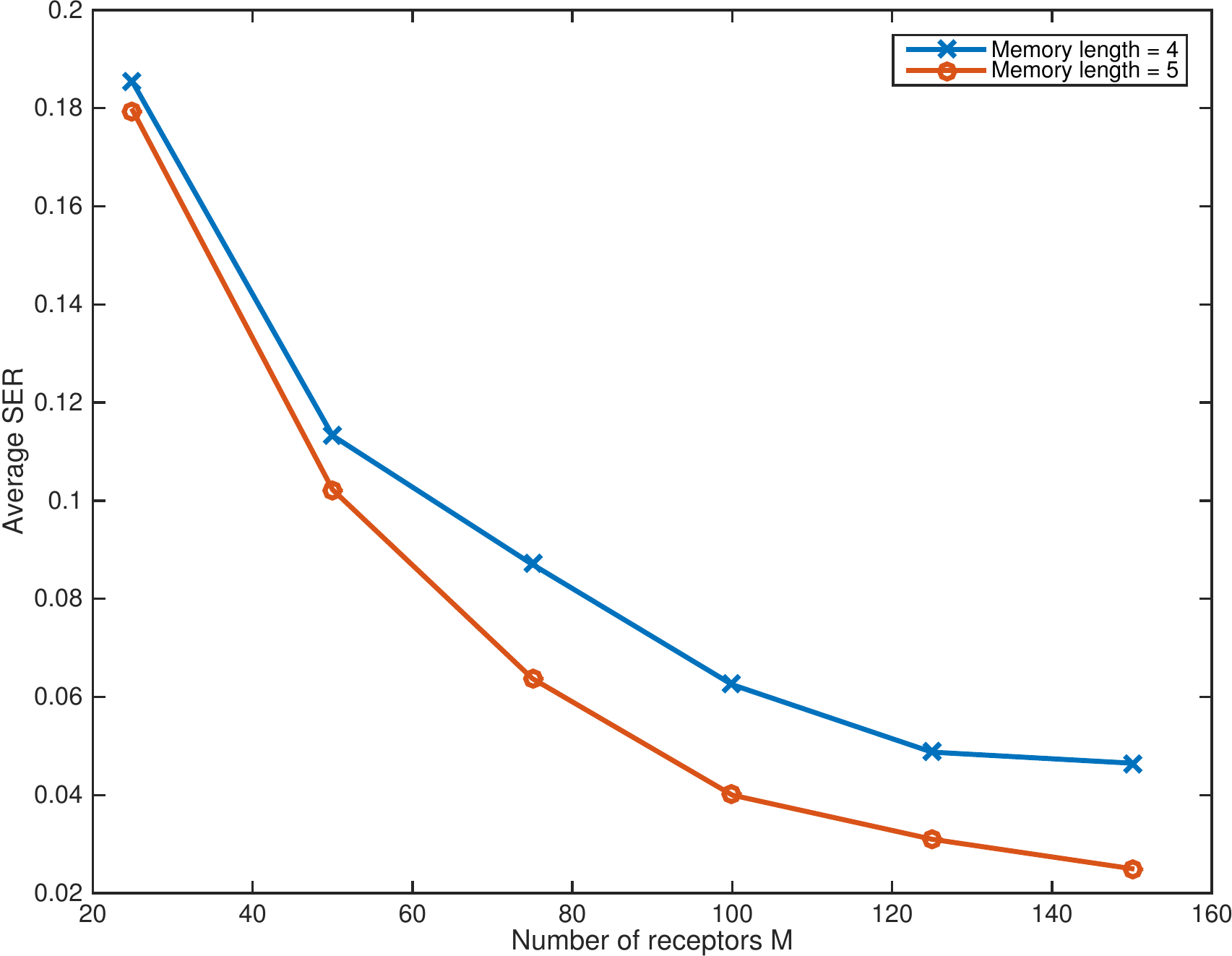}
\caption{The average SER versus number of receptors. The ISI case.}
\label{fig:isi}
\end{center}
\end{figure}

The aim of this experiment is to study the performance of the demodulator in the presence of ISI. We use the decision feedback method described in Section \ref{sec:isi} together with the approximation decomposition in \eqref{eqn:sigma_sum}. We vary the number of receptors from 25 to 150. We use two different memory lengths $\ell$ of 4 and 5. If the memory length is $\ell$, we express the mean number of output molecules at the current symbol interval as a sum of $\ell$ terms, i.e. a generalisation of \eqref{eqn:sigma_sum} to $\ell$ terms. We carry out the simulation for a duration of 20 symbol lengths and compute the average SER over 20 symbols. Figure \ref{fig:isi} shows the average SER versus the number of receptors. It can be seen that an increasing number of receptors can also be used to deal with ISI. 

\section{Conclusions and future work} 
\label{sec:con}
This paper studies a diffusion-based molecular communication network that uses different sets of chemical reactions to represent different transmission symbols. We focus on the demodulation problem. We assume the receiver uses a ligand-receptor binding process and uses the continuous history of the number of ligand-receptor complexes over time as the input signal to the demodulator. We derive the maximum a posteriori demodulator by solving a Bayesian filtering problem. 

\appendix

\ifsingle 
\section{Proof of Equation \eqref{eqn:predictb} } 
\label{app:proofA} 

Let $s$ denote the transmitted symbol, our aim is to determine $\mathbf{P}[b(t+\Delta t) | s, {\cal B}(t)]$ in terms of the quantity at time $t$. Recalling that $(N(t),b(t))$ is the state of the CTMP and since only ${\cal B}(t)$ is observed, the problem of predicting $b(t+\Delta t)$ from ${\cal B}(t)$ is a Bayesian filtering or hidden Markov model problem. The first step is to condition on the state of the system, as follows: 
\begin{align}
& \mathbf{P}[b(t+\Delta t) | s, {\cal B}(t)] \\
& =  \sum_{i} \mathbf{P}[N(t+\Delta t) = \eta_i, b(t+\Delta t) | s, {\cal B}(t)]  \\
& =  \sum_{i} \sum_{j} \mathbf{P}[N(t+\Delta t) = \eta_i, b(t+\Delta t) | s, N(t) = \eta_j, {\cal B}(t)] 
\mathbf{P}[N(t) = \eta_j | s,  {\cal B}(t)]  \label{eqn:app:int1} \\
& = \sum_{i} \sum_{j} \mathbf{P}[N(t+\Delta t) = \eta_i, b(t+\Delta t) | s, N(t) = \eta_j, b(t)]  \mathbf{P}[N(t) = \eta_j | s,  {\cal B}(t)] 
\label{eqn:app:condp}
\end{align}
where we have used the Markov property 
$\mathbf{P}[N(t+\Delta t) = \eta_i, b(t+\Delta t) | s, N(t) = \eta_j, {\cal B}(t)]  = \mathbf{P}[N(t+\Delta t) = \eta_i, b(t+\Delta t) | s, N(t) = \eta_j, b(t)]$ to arrive at Equation \eqref{eqn:app:condp}. 

We now focus on the term $\mathbf{P}[N(t+\Delta t) = \eta_i, b(t+\Delta t) | s, N(t) = \eta_j, b(t)]$ in Equation \eqref{eqn:app:condp}. This term is the state transition probability. Using the CTMP in Section \ref{sec:model},  we have
\begin{align}
& \mathbf{P}[N(t+\Delta t) = \eta_i, b(t+\Delta t) | s, N(t) = \eta_j, b(t)] \\ \nonumber 
=  & \delta_{b(t+\Delta t),b(t) + 1} P_1 + \delta_{b(t+\Delta t), b(t) - 1} P_2 + \delta_{b(t+\Delta t), b(t)} P_3 \label{eqn:app:p123} 
\end{align}
where
\begin{align}
P_1 & =   \delta_{\eta_i, \eta_j - \mathbb{1}_R} \lambda \eta_{j,R} (M - b(t)) \; \Delta t \\
P_2 & =    \delta_{\eta_i, \eta_j + \mathbb{1}_R} \mu b(t) \; \Delta t \\
P_3 & =   \eta_{i \rightarrow j} 
\end{align}
where $\eta_{j,R}$ is the $R$-th element of $\eta_j$, i.e. there are $\eta_{j,R}$ signalling molecules in the receiver voxel, and 
\begin{align}
\eta_{i \rightarrow j} = 
\left\{ 
\begin{array}{ll}
d_{ij} \; \Delta t & \mbox{ if } i \neq j \\
1 -  (\lambda \eta_{j,R} (M - b(t)) - \mu b(t) - d_{jj}) \; \Delta t & \mbox{ if } i = j
\end{array}
\right. 
\end{align}
where 
\begin{align}
d_{jj} = \sum_{i \neq j} d_{ij}
\end{align}


By substituting Equation \eqref{eqn:app:p123} into Equation \eqref{eqn:app:condp}, we have 
\begin{align}
\mathbf{P}[b(t+\Delta t) | s, {\cal B}(t)] = \delta_{b(t+\Delta t), b(t) + 1} Q_1 + \delta_{b(t+\Delta t), b(t) - 1}  Q_2 + \delta_{b(t+\Delta t), b(t)} Q_3 
\end{align} 
where 
\begin{align}
Q_{\ell} = \sum_{i} \sum_{j} P_{\ell} \mathbf{P}[N(t) = \eta_j | s,  {\cal B}(t)] 
\end{align}

We will now determine $Q_1$, $Q_2$ and $Q_3$. 

For $Q_1$, we have
\begin{align}
Q_1 
& = \sum_{i} \sum_{j} \delta_{\eta_i, \eta_j - \mathbb{1}_R} \lambda \eta_{j,R} (M - b(t)) \; \Delta t \; \mathbf{P}[N(t) = \eta_j | s,  {\cal B}(t)]
      \nonumber  \\
& = \lambda (M - b(t))  \; \Delta t  \;  \sum_{i} \sum_{j}  \delta_{\eta_i, \eta_j - \mathbb{1}_R} \eta_{j,R}  \mathbf{P}[N(t) = \eta_j | s,  {\cal B}(t)] 
      \nonumber \\
& =  \lambda (M - b(t))  \; \Delta t  \;  \sum_{j \; s.t. \; \eta_{j,R} \geq 1}  \eta_{j,R}  \mathbf{P}[N(t) = \eta_j | s,  {\cal B}(t)] 
       \label{eqn:app:q1:1} \\    
& =  \lambda (M - b(t))  \; \Delta t  \;  \sum_{j}  \eta_{j,R}  \mathbf{P}[N(t) = \eta_j | s,  {\cal B}(t)] 
       \label{eqn:app:q1:2} \\           
& =  \lambda (M - b(t))  \; \Delta t  \;  \mathbf{E} [n_R(t) | s,  {\cal B}(t)] 
\end{align}

Note that in Equation \eqref{eqn:app:q1:1}, the sum is over all states $\eta_i$ with at least one signalling molecule in the receiver voxel, i.e. $\eta_{j,R} \geq 1$. Since the summand in Equation \eqref{eqn:app:q1:1} is zero if $\eta_{j,R} = 0$, we get the same result if we are to sum over all possible states, that is why Equation \eqref{eqn:app:q1:2} holds.

For $Q_2$, we have
\begin{align}
Q_2 
& = \sum_{i} \sum_{j}   \delta_{\eta_i, \eta_j + \mathbb{1}_R} \mu b(t) \; \Delta t \; \mathbf{P}[N(t) = \eta_j | s,  {\cal B}(t)]  \nonumber \\
& =  \mu b(t) \; \Delta t
\sum_{i} \sum_{j}  \delta_{\eta_i, \eta_j + \mathbb{1}_R} \; \mathbf{P}[N(t) = \eta_j | s,  {\cal B}(t)]   \label{eqn:app:q2:1} \\
& =  \mu b(t) \; \Delta t
\sum_{j}  \mathbf{P}[N(t) = \eta_j | s,  {\cal B}(t)]   \label{eqn:app:q2:2} \\
& =  \mu b(t) \; \Delta t \; 
\end{align}

Note that Equation \eqref{eqn:app:q2:2} follows from Equation \eqref{eqn:app:q2:1} because for every $\eta_j$, there is a unique $\eta_i$ such that $\eta_i = \eta_j + \mathbb{1}_R$ holds. 

For $Q_3$, we have
\begin{align}
Q_3 
 =  &  \sum_{i} \sum_{j}  \eta_{i \rightarrow j} \mathbf{P}[N(t) = \eta_j | s,  {\cal B}(t)] \nonumber \\
  = & \sum_{i} \sum_{j \neq i}  (d_{ij} \; \Delta t) \mathbf{P}[N(t) = \eta_j | s,  {\cal B}(t)] + \nonumber \\
&  \sum_{j}  (1 - \lambda \eta_{j,R} (M - b(t)) \; \Delta t - \mu b(t) \; \Delta t - d_{jj} \; \Delta t)
 \mathbf{P}[N(t) = \eta_j | s,  {\cal B}(t)]  \nonumber \\
   =  &
     \sum_{j} (1 - \lambda \eta_{j,R} (M - b(t)) \; \Delta t - \mu b(t) \; \Delta t)  \mathbf{P}[N(t) = \eta_j | s,  {\cal B}(t)] + \nonumber \\
& 
(\sum_{i} \sum_{j \neq i} d_{ij} \mathbf{P}[N(t) = \eta_j | s,  {\cal B}(t)]  - \sum_{j} d_{jj} \mathbf{P}[N(t) = \eta_j | s,  {\cal B}(t)]) \; \Delta t \nonumber \\
 = &  (1 - \lambda (M - b(t)) {\mathbf E}[n_R(t) | s,  {\cal B}(t)] \; \Delta t - \mu b(t)  \; \Delta t) + \nonumber \\
& 
\underbrace{(\sum_{i} \sum_{j \neq i} d_{ij} \mathbf{P}[N(t) = \eta_j | s,  {\cal B}(t)]  - \sum_{j} \sum_{i \neq j} d_{ij} \mathbf{P}[N(t) = \eta_j | s,  {\cal B}(t)])}_{= 0} 
\; \Delta t \nonumber \\
 = &  (1 - \lambda (M - b(t)) {\mathbf E}[n_R(t) | s,  {\cal B}(t)] \; \Delta t - \mu b(t)  \; \Delta t)
\end{align} 

Having obtained $Q_1$, $Q_2$ and $Q_3$, we arrive at: 
\begin{align}
 \mathbf{P}[b(t+\Delta t) | s, {\cal B}(t)] 
 =  & \delta_{b(t+\Delta t), b(t) + 1} \lambda (M - b(t))  \; \Delta t  \;  \mathbf{E} [n_R(t) | s,  {\cal B}(t)]  +  \nonumber \\ 
 & \delta_{b(t+\Delta t), b(t) - 1} \mu b(t) \; \Delta t \; + \nonumber  \\
& \delta_{b(t+\Delta t), b(t)} (1 - \lambda (M - b(t)) {\mathbf E}[n_R(t) | s,  {\cal B}(t)] \; \Delta t - \mu b(t)  \; \Delta t) \label{eqn:app:predictb}
\end{align}
Note that Equation \eqref{eqn:app:predictb} is the same as Equation \eqref{eqn:predictb} in the main text. 

\section{Proof of Equation \eqref{eqn:logpp_dd}} 
\label{app:proofB}
From Equation \eqref{eqn:logpp_prelim}, we have: 
\begin{align}
\frac{d L_s(t)}{dt} = \lim_{\Delta t \rightarrow 0} \frac{\log (\mathbf{P}[b(t+\Delta t) | s, {\cal B}(t)])}{\Delta t} - 
\lim_{\Delta t \rightarrow 0} \frac{\log (\mathbf{P}[b(t+\Delta t) | {\cal B}(t)])}{\Delta t}
\label{eqn:logpp:app:prelim}
\end{align}
Note that the second term on the RHS is independent of transmission symbol $s$, we will focus on the first term. 

Note that $\mathbf{P}[b(t+\Delta t) | s, {\cal B}(t)]$, which is given in Equation \eqref{eqn:app:predictb}, is a sum three terms with multipliers $\delta_{b(t+\Delta t), b(t) + 1}$, $\delta_{b(t+\Delta t), b(t) - 1}$ and $\delta_{b(t+\Delta t), b(t)}$. Since these multipliers are mutually exclusive, we have:
\begin{align}
\log \left( \mathbf{P}[b(t+\Delta t) | s, {\cal B}(t)] \right)  
= & 
\delta_{b(t+\Delta t), b(t) + 1} \log \left( \lambda (M - b(t))  \; \Delta t  \;  \mathbf{E} [n_R(t) | s,  {\cal B}(t)]   \right) + \nonumber \\
& \delta_{b(t+\Delta t), b(t) - 1} \log \left( \mu b(t) \; \Delta t   \right) \nonumber + \\
& \delta_{b(t+\Delta t), b(t)} \log \left( (1 - \lambda (M - b(t)) {\mathbf E}[n_R(t) | 1,  {\cal B}(t)] \; \Delta t - \mu b(t)  \; \Delta t)   \right) 
\label{app:logapprox}  \\
 \approx & 
\delta_{b(t+\Delta t), b(t) + 1} \log \left( \mathbf{E} [n_R(t) | s,  {\cal B}(t)] \right) - \nonumber \\ 
& \delta_{b(t+\Delta t), b(t)} \lambda (M - b(t)) {\mathbf E}[n_R(t) | s,  {\cal B}(t)] \; \Delta t + \nonumber \\ 
& \tilde{P}(t)  
\label{eqn:app:likelihood} 
\end{align}
where we have used the approximation $\log(1 + \alpha \; \Delta t) \approx \alpha \; \Delta t$ in the last term of Equation \eqref{app:logapprox} 
and have collected all terms that do not depend on $s$ in $\tilde{P}(t)$.  

By substituting Equation \eqref{eqn:app:likelihood} into Equation \eqref{eqn:logpp:app:prelim}, and taking limit $\Delta t \rightarrow 0$, we have
\begin{align}
\frac{dL_s(t)}{dt} 
= & \lim_{\Delta t \rightarrow 0} \frac{\delta_{b(t+\Delta t), b(t) + 1} }{\Delta t}
\log \left( \mathbf{E} [n_R(t) | s,  {\cal B}(t)]  \right) - \nonumber \\
& \delta_{b(t+\Delta t), b(t)} \lambda (M - b(t))  
\left( {\mathbf E}[n_R(t) | s,  {\cal B}(t)] ]  \right) + \tilde{L}(t) \label{eqn:app:logmapm1}  \\
= & \frac{dU(t)}{dt} 
\log \left( \mathbf{E} [n_R(t) | s,  {\cal B}(t)]  \right) - 
\lambda (M - b(t))  
\left( {\mathbf E}[n_R(t) | s,  {\cal B}(t)]  \right) + \tilde{L}(t) \label{eqn:app:logmap}
\end{align}
where all terms that are independent of $s$ have been collected in $\tilde{L}(t)$. Note that $\tilde{L}(t)$ contains some terms that diverges but this is not an issue because for demodulation it is their relative difference $L_{s_1}(t) - L_{s_2}(t)$ (for any two symbols $s_1$ and $s_2$) that matters. 

Note also that we have used the following reasonings to arrive at Equation \eqref{eqn:app:logmap} from Equation \eqref{eqn:app:logmapm1}:
\begin{enumerate}
\item The term $\lim_{\Delta t \rightarrow 0} \frac{\delta_{b(t+\Delta t), b(t) + 1} }{\Delta t}$ is an impulse whenever a receptor changes from the unbound to the bound state. This is precisely $\frac{dU(t)}{dt} $. 
\item The term $\delta_{b(t+\Delta t), b(t)}$ is only zero when the number of bound receptors changes and the number of such changes is finite. In other words, $\delta_{b(t+\Delta t), b(t)}=1$ with probability one. This allows us to drop $\delta_{b(t+\Delta t), b(t)}$. 
\end{enumerate} 

Finally, note that Equation \eqref{eqn:app:logmap} is the same as Equation \eqref{eqn:logpp_dd}.

\else 
\section{Proof of Equation \eqref{eqn:predictb} } 
\label{app:proofA} 

Let $s$ denote the transmitted symbol, our aim is to determine $\mathbf{P}[b(t+\Delta t) | s, {\cal B}(t)]$ in terms of the quantity at time $t$. Recalling that $(N(t),b(t))$ is the state of the CTMP and since only ${\cal B}(t)$ is observed, the problem of predicting $b(t+\Delta t)$ from ${\cal B}(t)$ is a Bayesian filtering or hidden Markov model problem. The first step is to condition on the state of the system, as follows: 
\begin{align}
& \mathbf{P}[b(t+\Delta t) | s, {\cal B}(t)] \\
 =  & \sum_{i} \mathbf{P}[N(t+\Delta t) = \eta_i, b(t+\Delta t) | s, {\cal B}(t)]  \\
 =  & \sum_{i} \sum_{j} \mathbf{P}[N(t+\Delta t) = \eta_i, b(t+\Delta t) | s, N(t) = \eta_j, {\cal B}(t)] \times \nonumber \\
& \mathbf{P}[N(t) = \eta_j | s,  {\cal B}(t)]  \label{eqn:app:int1} \\
= &  \sum_{i} \sum_{j} \mathbf{P}[N(t+\Delta t) = \eta_i, b(t+\Delta t) | s, N(t) = \eta_j, b(t)] \times \nonumber \\
&  \mathbf{P}[N(t) = \eta_j | s,  {\cal B}(t)] 
\label{eqn:app:condp}
\end{align}
where we have used the Markov property 
$\mathbf{P}[N(t+\Delta t) = \eta_i, b(t+\Delta t) | s, N(t) = \eta_j, {\cal B}(t)]  = \mathbf{P}[N(t+\Delta t) = \eta_i, b(t+\Delta t) | s, N(t) = \eta_j, b(t)]$ to arrive at Equation \eqref{eqn:app:condp}. 

We now focus on the term $\mathbf{P}[N(t+\Delta t) = \eta_i, b(t+\Delta t) | s, N(t) = \eta_j, b(t)]$ in Equation \eqref{eqn:app:condp}. This term is the state transition probability. Using the CTMP in Section \ref{sec:model},  we have
\begin{align}
& \mathbf{P}[N(t+\Delta t) = \eta_i, b(t+\Delta t) | s, N(t) = \eta_j, b(t)]  \nonumber  \\ 
=  & \delta_{b(t+\Delta t), b(t) + 1} P_1 + \delta_{b(t+\Delta t), b(t) - 1} P_2 +  \nonumber  \\ 
&  \delta_{b(t+\Delta t),b(t)} P_3 \label{eqn:app:p123} 
\end{align}
where
\begin{align}
P_1 & =   \delta_{\eta_i, \eta_j - \mathbb{1}_R} \lambda \eta_{j,R} (M - b(t)) \; \Delta t \\
P_2 & =    \delta{\eta_i, \eta_j + \mathbb{1}_R} \mu b(t) \; \Delta t \\
P_3 & =   \eta_{i \rightarrow j} 
\end{align}
where $\eta_{j,R}$ is the $R$-th element of $\eta_j$, i.e. there are $\eta_{j,R}$ signalling molecules in the receiver voxel, and 
\begin{align}
\eta_{i \rightarrow j} = 
\left\{ 
\begin{array}{ll}
d_{ij} \Delta t & \mbox{if } i \neq j \\
1 -  (\lambda \eta_{j,R} (M - b(t)) - \mu b(t) - d_{jj}) \Delta t & \mbox{if } i = j
\end{array}
\right. 
\end{align}
where 
\begin{align}
d_{jj} = \sum_{i \neq j} d_{ij}
\end{align}


By substituting Equation \eqref{eqn:app:p123} into Equation \eqref{eqn:app:condp}, we have 
\begin{align}
& \mathbf{P}[b(t+\Delta t) | s, {\cal B}(t)] = \nonumber \\
& \delta_{b(t+\Delta t), b(t) + 1} Q_1 + \delta_{b(t+\Delta t), b(t) - 1}  Q_2 + \nonumber \\
& \delta_{b(t+\Delta t), b(t)} Q_3 
\end{align} 
where 
\begin{align}
Q_{\ell} = \sum_{i} \sum_{j} P_{\ell} \mathbf{P}[N(t) = \eta_j | s,  {\cal B}(t)] 
\end{align}

We will now determine $Q_1$, $Q_2$ and $Q_3$. 

For $Q_1$, we have
\begin{align}
Q_1 
= &  \sum_{i} \sum_{j} \delta_{\eta_i, \eta_j - \mathbb{1}_R} \lambda \eta_{j,R} (M - b(t)) \; \Delta t \; \times  \nonumber \\
& \mathbf{P}[N(t) = \eta_j | s,  {\cal B}(t)]
      \nonumber  \\
= & \lambda (M - b(t))  \; \Delta t  \;  \; \times  \nonumber \\
  &  \sum_{i} \sum_{j}  \delta_{\eta_i, \eta_j - \mathbb{1}_R} \eta_{j,R}  \mathbf{P}[N(t) = \eta_j | s,  {\cal B}(t)] 
      \nonumber \\
 = & \lambda (M - b(t))  \; \Delta t  \;  \sum_{j \; s.t. \; \eta_{j,R} \geq 1}  \eta_{j,R}  \mathbf{P}[N(t) = \eta_j | s,  {\cal B}(t)] 
       \label{eqn:app:q1:1} \\    
 =  & \lambda (M - b(t))  \; \Delta t  \;  \sum_{j}  \eta_{j,R}  \mathbf{P}[N(t) = \eta_j | s,  {\cal B}(t)] 
       \label{eqn:app:q1:2} \\           
 =  & \lambda (M - b(t))  \; \Delta t  \;  \mathbf{E} [n_R(t) | s,  {\cal B}(t)] 
\end{align}
Note that in Equation \eqref{eqn:app:q1:1}, the sum is over all states $\eta_i$ with at least one signalling molecule in the receiver voxel, i.e. $\eta_{j,R} \geq 1$. Since the summand in Equation \eqref{eqn:app:q1:1} is zero if $\eta_{j,R} = 0$, we get the same result if we are to sum over all possible states, that is why Equation \eqref{eqn:app:q1:2} holds.

For $Q_2$, we have
\begin{align}
Q_2 
& = \sum_{i} \sum_{j}   \delta_{\eta_i, \eta_j + \mathbb{1}_R} \mu b(t) \; \Delta t \; \mathbf{P}[N(t) = \eta_j | s,  {\cal B}(t)]  \nonumber \\
& =  \mu b(t) \; \Delta t
\sum_{i} \sum_{j}  \delta_{\eta_i , \eta_j + \mathbb{1}_R} \; \mathbf{P}[N(t) = \eta_j | s,  {\cal B}(t)]   \label{eqn:app:q2:1} \\
& =  \mu b(t) \; \Delta t
\sum_{j}  \mathbf{P}[N(t) = \eta_j | s,  {\cal B}(t)]   \label{eqn:app:q2:2} \\
& =  \mu b(t) \; \Delta t \; 
\end{align}
Note that Equation \eqref{eqn:app:q2:2} follows from Equation \eqref{eqn:app:q2:1} because for every $\eta_j$, there is a unique $\eta_i$ such that $\eta_i = \eta_j + \mathbb{1}_R$ holds. 

For $Q_3$, we have
\begin{align}
& Q_3  \nonumber \\
 =  &  \sum_{i} \sum_{j}  \eta_{i \rightarrow j} \mathbf{P}[N(t) = \eta_j | s,  {\cal B}(t)] \nonumber \\
  = & \sum_{i} \sum_{j \neq i}  (d_{ij} \; \Delta t) \mathbf{P}[N(t) = \eta_j | s,  {\cal B}(t)] + \nonumber \\
&  \sum_{j}  (1 - \lambda \eta_{j,R} (M - b(t)) \; \Delta t - \mu b(t) \; \Delta t - d_{jj} \; \Delta t) \times \nonumber \\
 & \mathbf{P}[N(t) = \eta_j | s,  {\cal B}(t)]  \nonumber \\
   =  &
     \sum_{j} (1 - \lambda \eta_{j,R} (M - b(t)) \; \Delta t - \mu b(t) \; \Delta t)   
      \mathbf{P}[N(t) = \eta_j | s,  {\cal B}(t)] + \nonumber \\
& 
(\sum_{i} \sum_{j \neq i} d_{ij} \mathbf{P}[N(t) = \eta_j | s,  {\cal B}(t)]  - \sum_{j} d_{jj} \mathbf{P}[N(t) = \eta_j | s,  {\cal B}(t)]) \; \Delta t \nonumber \\
 = &  (1 - \lambda (M - b(t)) {\mathbf E}[n_R(t) | s,  {\cal B}(t)] \; \Delta t - \mu b(t)  \; \Delta t) + (\Delta t) \;  \times \nonumber \\
& 
\underbrace{(\sum_{i} \sum_{j \neq i} d_{ij} \mathbf{P}[N(t) = \eta_j | s,  {\cal B}(t)]  - \sum_{j} \sum_{i \neq j} d_{ij} \mathbf{P}[N(t) = \eta_j | s,  {\cal B}(t)])}_{= 0} 
 \nonumber \\
 = &  (1 - \lambda (M - b(t)) {\mathbf E}[n_R(t) | s,  {\cal B}(t)] \; \Delta t - \mu b(t)  \; \Delta t)
\end{align} 

Having obtained $Q_1$, $Q_2$ and $Q_3$, we arrive at: 
\begin{align}
&  \mathbf{P}[b(t+\Delta t) | s, {\cal B}(t)] \nonumber \\ 
 =  & \delta_{b(t+\Delta t), b(t) + 1} \lambda (M - b(t))  \; \Delta t  \;  \mathbf{E} [n_R(t) | s,  {\cal B}(t)]    \nonumber \\ 
 & + \delta_{b(t+\Delta t), b(t) - 1} \mu b(t) \; \Delta t \; + \nonumber  \\
& \delta_{b(t+\Delta t), b(t)} \times \nonumber \\ 
& (1 - \lambda (M - b(t)) {\mathbf E}[n_R(t) | s,  {\cal B}(t)] \; \Delta t - \mu b(t)  \; \Delta t) \label{eqn:app:predictb}
\end{align}
Note that Equation \eqref{eqn:app:predictb} is the same as Equation \eqref{eqn:predictb} in the main text. 

\section{Proof of Equation \eqref{eqn:logpp_dd}} 
\label{app:proofB}
From Equation \eqref{eqn:logpp_prelim}, we have: 
\begin{align}
\frac{d L_s(t)}{dt} = &  \lim_{\Delta t \rightarrow 0} \frac{\log (\mathbf{P}[b(t+\Delta t) | s, {\cal B}(t)])}{\Delta t} - \nonumber \\
& \lim_{\Delta t \rightarrow 0} \frac{\log (\mathbf{P}[b(t+\Delta t) | {\cal B}(t)])}{\Delta t}
\label{eqn:logpp:app:prelim}
\end{align}
Note that the second term on the RHS is independent of transmission symbol $s$, we will focus on the first term. 

Note that $\mathbf{P}[b(t+\Delta t) | s, {\cal B}(t)]$, which is given in Equation \eqref{eqn:app:predictb}, is a sum three terms with multipliers $\delta_{b(t+\Delta t), b(t) + 1}$, $\delta_{b(t+\Delta t), b(t) - 1}$ and $\delta_{b(t+\Delta t), b(t)}$. Since these multipliers are mutually exclusive, we have:
\begin{align}
& \log \left( \mathbf{P}[b(t+\Delta t) | s, {\cal B}(t)] \right)  \nonumber  \\
= &  
\delta_{b(t+\Delta t) , b(t) + 1} \times \nonumber \\ 
& \log \left( \lambda (M - b(t))  \; \Delta t  \;  \mathbf{E} [n_R(t) | s,  {\cal B}(t)]   \right) + \nonumber \\
& \delta_{b(t+\Delta t) , b(t) - 1} \log \left( \mu b(t) \; \Delta t   \right) \nonumber + \\
& \delta_{b(t+\Delta t) , b(t)} \times \nonumber \\ 
&\log \left( (1 - \lambda (M - b(t)) {\mathbf E}[n_R(t) | 1,  {\cal B}(t)] \; \Delta t - \mu b(t)  \; \Delta t)   \right) 
\label{app:logapprox}  \\
 \approx & 
\delta_{b(t+\Delta t) , b(t) + 1} \log \left( \mathbf{E} [n_R(t) | s,  {\cal B}(t)] \right) - \nonumber \\ 
& \delta_{b(t+\Delta t) , b(t)} \lambda (M - b(t)) {\mathbf E}[n_R(t) | s,  {\cal B}(t)] \; \Delta t + \nonumber \\ 
& \tilde{P}(t)  
\label{eqn:app:likelihood} 
\end{align}
where we have used the approximation $\log(1 + \alpha \; \Delta t) \approx \alpha \; \Delta t$ in the last term of Equation \eqref{app:logapprox} 
and have collected all terms that do not depend on $s$ in $\tilde{P}(t)$.  

By substituting Equation \eqref{eqn:app:likelihood} into Equation \eqref{eqn:logpp:app:prelim}, and taking limit $\Delta t \rightarrow 0$, we have
\begin{align}
& \frac{dL_s(t)}{dt} \nonumber \\
= & \lim_{\Delta t \rightarrow 0} \frac{\delta_{b(t+\Delta t) , b(t) + 1} }{\Delta t}
\log \left( \mathbf{E} [n_R(t) | s,  {\cal B}(t)]  \right) - \nonumber \\
& \delta_{b(t+\Delta t) , b(t)} \lambda (M - b(t))  
\left( {\mathbf E}[n_R(t) | s,  {\cal B}(t)] ]  \right) + \tilde{L}(t) \label{eqn:app:logmapm1}  \\
= & \frac{dU(t)}{dt} 
\log \left( \mathbf{E} [n_R(t) | s,  {\cal B}(t)]  \right) -   \nonumber \\
& \lambda (M - b(t))  
\left( {\mathbf E}[n_R(t) | s,  {\cal B}(t)]  \right) + \tilde{L}(t) \label{eqn:app:logmap}
\end{align}
where all terms that are independent of $s$ have been collected in $\tilde{L}(t)$. Note that $\tilde{L}(t)$ contains some terms that diverges but this is not an issue because for demodulation it is their relative difference $L_{s_1}(t) - L_{s_2}(t)$ (for any two symbols $s_1$ and $s_2$) that matters. 

Note also that we have used the following reasonings to arrive at Equation \eqref{eqn:app:logmap} from Equation \eqref{eqn:app:logmapm1}:
\begin{enumerate}
\item The term $\lim_{\Delta t \rightarrow 0} \frac{\delta_{b(t+\Delta t) , b(t) + 1} }{\Delta t}$ is an impulse whenever a receptor changes from the unbind to the bind state. This is precisely $\frac{dU(t)}{dt} $. 
\item The term $\delta_{b(t+\Delta t) , b(t)}$ is only zero when the number of bound receptors changes and the number of such changes is finite. In other words, $\delta_{b(t+\Delta t) , b(t)}=1$ with probability one. This allows us to drop $\delta_{b(t+\Delta t) = b(t)}$. 
\end{enumerate} 

Finally, note that Equation \eqref{eqn:app:logmap} is the same as Equation \eqref{eqn:logpp_dd}. 

\fi 

\section{An example end-to-end model}
\label{app:e2e_ex} 
For this example, we assume the transmission medium consists of 3 voxels as illustrated in Figure \ref{fig:small}. The transmitter and receiver are assumed to be located in, respectively, Voxels 1 and 3. We assume reflecting boundary condition which means the signalling molecules cannot leave the medium. 

This Appendix presents an example end-to-end model. For this end-to-end model, the transmitter is assumed to send Symbol 0 which means it uses the set of chemical reactions corresponding to this symbol. We therefore view a transmitter as a set of chemical reactions located within the transmitter voxel. It is still an open problem what chemical reactions are good for communication performance. The example being used here is not meant to promote the use of a particular set of chemical reactions but our purpose is to show how a set of chemical reactions can be modelled by a CTMP.

For this example, we assume that the production of the signalling molecules $S$ requires two intermediate chemical species $F$ and $G$, which are produced by RNA${}_1$ and RNA${}_2$. There are four reactions and they are assumed to take place within the transmitter voxel only. The four chemical reactions are: 
\begin{align}
\cee{
RNA_1 &  ->[k_1] RNA_1 + F
} 
\label{cr:t:r1}  \\
\cee{
RNA_2 & ->[k_2] RNA_2 + G
} 
\label{cr:t:r2}  \\
\cee{
F  & ->[k_3] S
} 
\label{cr:t:r3}  
\\
\cee{
S + G &  ->[k_4] \phi
} 
\label{cr:t:r4}
\end{align}
Reaction \eqref{cr:t:r1} says that the molecules of $F$ are produced at a mean rate of $k_1$. Similarly, according to Reaction \eqref{cr:t:r2}, $G$ is produced at a mean rate of $k_2$. Reaction \eqref{cr:t:r3} says that $F$ is converted to $S$ at a mean rate equals to $k_3$ times the number of $F$ molecules in the transmitter voxel. Reaction \eqref{cr:t:r4} says that $S$ and $G$ can react to produce a molecule $\phi$ that we are not interested to keep track of in the mathematical model. If an $S$ (or a $G$) molecule takes part in Reaction \eqref{cr:t:r4}, we can consider this $S$ ($G$) molecule has left the system permanently after the reaction. The rate of Reaction \eqref{cr:t:r4} is $k_4$ times the number of $G$ molecules and the number of signalling molecules $S$ in the transmitter voxel. 

We assume that the chemical species RNA${}_1$, RNA${}_2$, $F$ and $G$ are found in the transmitter voxel only, and they cannot leave the transmitter voxel. This means that we do not need to consider the diffusion of these chemical species. The only diffusible chemical species in the entire system is the signalling molecule $S$. We also assume that there is only one of each RNA${}_1$ and RNA${}_2$ and their counts remain constant.

In order to define the state of the system, we make the following definitions: $n_i(t)$ is the number of signalling molecules in Voxel $i$ at time $t$, $n_F(t)$ and $n_G(t)$ are respectively the number of $F$ and $G$ molecules at time $t$, $n_A(t)$ is the cumulative number of molecules that have left the system at time $t$ and $b(t)$ is the number of complexes (or bound receptors) at time $t$. Since a receptor can either be unbound or in a complex, the number of unbound receptors at time $t$ is $M-b(t)$; therefore, the mathematical model only has to keep track of either the number of unbound receptors or the number of complexes, and we have chosen to keep track of the latter. The state of the system is completely specified by these seven molecular counts: $n_1(t),n_2(t),n_3(t),n_F(t),n_G(t),n_A(t)$ and $b(t)$. All the molecular counts should be non-negative integers (i.e. belonging to the set ${\mathbb Z}_{\ge 0}$) and a further restriction is that $0 \leq b(t) \leq M$ or we write $b(t) \in {\mathbb Z}_{[0,M]}$. 

We define the vector $N(t)$ as 
\begin{align}
N(t) = \left[ \begin{array}{cccccc}
n_{1}(t) & n_2(t) & n_{3}(t)  & n_{F}(t) & n_{G}(t) & n_{A}(t) 
\end{array}
\right]^T
\label{eqn:ex:state}
\end{align} 
where the superscript $^{T}$ is used to denote matrix transpose. 

Based on the definition of $N(t)$, the state of the system is the tuple $(N(t),b(t))$ and a valid state must be an element of the set ${\cal S} =  \mathbb{Z}_{\ge 0}^6 \times \mathbb{Z}_{[0,M]}$. The state of the system changes when a reaction or diffusion event occurs. Our modelling assumptions mean that reactions can only take place in the transmitter or the receiver voxels. The reactions in the transmitter voxel are \eqref{cr:t:r1}$-$\eqref{cr:t:r4}. The reactions taking place in the receiver voxel are \eqref{cr:on} and \eqref{cr:off}. The only diffusible chemical species in this system is the signalling molecule $S$. Within an infinitesimal time $\Delta t$, {\bf at most one} diffusion or reaction event can occur. Therefore, the dynamics of the system can be specified by the transition probability from state $(N(t),b(t))$ to $(N(t + \Delta t),b(t + \Delta t))$. We will now specify these transition probabilities and we begin with the transmitter. 

Four possible reaction events \eqref{cr:t:r1}--\eqref{cr:t:r4} can take place in the transmitter voxel. An occurrence of Reaction \eqref{cr:t:r1} increases the number of $F$ molecules in the transmitter voxel by 1 and this occurs at a mean rate of $k_1$. By defining $\mathbb{1}_i$ to be the standard basis vector with a `1' at the $i$-th position, we can write the state transition probability due to Reaction \eqref{cr:t:r1} as: 
\begin{align}
{
{\bf P}[N(t+\Delta t) = N(t)+\mathbb{1}_4, b(t+\Delta t) = b(t) | N(t), b(t)] = k_1 \; \Delta t
}
\label{eqn:ex:t1}
\end{align}
Note that we have used $\mathbb{1}_4$ because $n_F(t)$ is increased by 1 if Reaction \eqref{cr:t:r1} occurs and $n_F(t)$ is the fourth element of $N(t)$ in the definition of $N(t)$ in \eqref{eqn:ex:state}. The right-hand side (RHS) of Equation \eqref{eqn:ex:t1} is the transition probability that Reaction \eqref{cr:t:r1} occurs in $(t,t+\Delta t)$, which is given by the reaction rate $k_1$ times $\Delta t$. 

We can write the transition probabilities due to Reactions \eqref{cr:t:r2}$-$\eqref{cr:t:r4} as:
\begin{align}
 {\bf P}[N(t+\Delta t) = N(t)+\mathbb{1}_5, b(t+\Delta t) = b(t) | N(t), b(t)] 
 &= k_2 \; \Delta t 
\label{eqn:ex:t2} \\
 {\bf P}[N(t+\Delta t) = N(t)-\mathbb{1}_4+\mathbb{1}_1, b(t+\Delta t) = b(t) | N(t), b(t)] 
 &= k_3 \; n_F(t) \; \Delta t
\label{eqn:ex:t3} \\
 {\bf P}[N(t+\Delta t) = N(t)-\mathbb{1}_5-\mathbb{1}_1+2\mathbb{1}_6, b(t+\Delta t) = b(t) | N(t), b(t)] 
& = k_4 \; n_G(t) \; n_1(t) \Delta t
\label{eqn:ex:t4} 
\end{align}
The rationale behind Equation \eqref{eqn:ex:t2} is similar to that of \eqref{eqn:ex:t1}. Equation \eqref{eqn:ex:t3} models Reaction \eqref{cr:t:r3}. If Reaction \eqref{cr:t:r3} occurs, an $F$ molecule is converted to an $S$ molecule, so the number of $F$ molecules $n_F(t)$ (which is the fourth element of $N(t)$) is decreased by 1 and the number of signalling molecule in the transmitter voxel $n_1(t)$ (which is the first element of $N(t)$) is increased by 1; this change in the number of molecules as a result of Reaction \eqref{cr:t:r3} can be written as $N(t+\Delta t) = N(t)-\mathbb{1}_4+\mathbb{1}_1$ in \eqref{eqn:ex:t3}. Equation \eqref{eqn:ex:t4} models Reaction \eqref{cr:t:r4}. When Reaction \eqref{cr:t:r4} occurs, a $G$ and an $S$ molecule in the transmitter are consumed, hence $-\mathbb{1}_5-\mathbb{1}_1$ in \eqref{eqn:ex:t4}. We are not interested to keep track of the molecules as a result of this reaction, we consider these two molecules have left the system permanently and add `2' to $n_A(t)$ which is at the sixth position of $N(t)$. The letter `A' here comes from 'absorbing' because once a molecule is added to $n_A(t)$, it will not leave. Note that the RHSs of \eqref{eqn:ex:t1}--\eqref{eqn:ex:t4} show the transition probabilities and they are of the form of the transition rate times $\Delta t$.

The state of the system can also be changed by signalling molecules diffusing from one voxel to another. For this example, there are four possible diffusion events, which take place when a signalling molecule diffuses from a voxel to its neighbouring voxel. The four diffusion events are: from Voxel 1 to Voxel 2, from Voxel 2 to Voxel 1, from Voxel 2 to Voxel 3, and from Voxel 3 to Voxel 2. The transition probabilities of these four events are:
\begin{align}
{\bf P}[N(t+\Delta t) = N(t)-\mathbb{1}_1+\mathbb{1}_2, b(t+\Delta t) = b(t) | N(t), b(t)] 
&
= d \; n_1(t) \; \Delta t
\label{eqn:ex:d1} \\
 {\bf P}[N(t+\Delta t) = N(t)+\mathbb{1}_1-\mathbb{1}_2, b(t+\Delta t) = b(t) | N(t), b(t)]  
& = d \; n_2(t) \; \Delta t
\label{eqn:ex:d2} \\
 {\bf P}[N(t+\Delta t) = N(t)-\mathbb{1}_2+\mathbb{1}_3, b(t+\Delta t) = b(t) | N(t), b(t)] 
& = d \; n_2(t) \; \Delta t
\label{eqn:ex:d3} \\
 {\bf P}[N(t+\Delta t) = N(t)+\mathbb{1}_2-\mathbb{1}_3, b(t+\Delta t) = b(t) | N(t), b(t)]  
&= d \; n_3(t) \; \Delta t
\label{eqn:ex:d4} 
\end{align}
Equation \eqref{eqn:ex:d1} is the probability that a signalling molecules diffuses from Voxel 1 to Voxel 2. The occurrence of this event means the number of signalling molecules in Voxel 1 ($= n_1(t)$, which is the first element of $N(t)$) is decreased by 1 while the number of signalling molecules in Voxel 2  ($= n_2(t)$, which is the second element of $N(t)$) is increased by 1. The probability of this occurring is $d \; \Delta t$. The explanation for the other three transition probabilities are similar. 

The last category of state transitions occurs when a receptor is bound or unbound according to chemical reactions \eqref{cr:on} and \eqref{cr:off}. The state transition probabilities are:
\begin{align}
 {\bf P}[N(t+\Delta t) = N(t)-\mathbb{1}_3, b(t+\Delta t) = b(t) + 1 | N(t), b(t)]  
 & = \lambda \; n_3(t) \; (M-b(t)) \; \Delta t
\label{eqn:ex:r1} \\
 {\bf P}[N(t+\Delta t) = N(t)+\mathbb{1}_3, b(t+\Delta t) = b(t) - 1 | N(t), b(t)]  
&= \mu \; b(t) \; \Delta t
\label{eqn:ex:r2} 
\end{align}
Equation \eqref{eqn:ex:r1} is the transition probability for receptor binding or the formation of new complex. This event occurs when a signalling molecule in the receiver voxel reacts with a unbound receptor to form a complex. As a result of this reaction, the number of signalling molecules in the receiver voxel (which is Voxel 3 in this example) is decreased by 1 and the number of complexes $b(t)$ is increased by 1. The rate of this event is proportional to the product of the number of signalling molecules in the receiver voxel $n_3(t)$ and the number of unbound receptors $(M-b(t))$. Equation \eqref{eqn:ex:r2} is the transition probability for a receptor to unbind. The unbinding reaction causes the number of signalling molecules in the receiver voxel $n_3(t)$ to increase by 1 while the number of complexes $b(t)$ to decrease by 1. The rate of this reaction is proportional to number of complexes $b(t)$. 

Equations \eqref{eqn:ex:t1} to \eqref{eqn:ex:r2} give the transition probabilities of the possible events that can occur when the state of the system is $(N(t),b(t))$. It is possible that no transitions occurs in the time interval $(t,t+\Delta t)$, the probability of this occurring is given by the complementary to that of an event occurring, that is, one minus the sum of the RHSs of Equations \eqref{eqn:ex:t1} to \eqref{eqn:ex:r2}. This completes the model.

\section{Difference between ${\mathbf E}[n_R(t) | s, {\cal B}(t)]$ and $\sigma_s(t)$} 
\label{app:C}
In Section \ref{sec:MAP}, we propose to replace ${\mathbf E}[n_R(t) | s, {\cal B}(t)]$ by $\sigma_s(t) = {\mathbf E}[n_R(t) | s]$. The difference between these two quantities is hard to compute because ligand-receptor binding is a non-linear process. In this Appendix, we will provide some justification of this replacement by considering the filtering problem of LTI systems. Consider the following continuous-time LTI system in state space form \cite{Franklin:cont}:
\begin{align}
\frac{dx(t)}{dt} =& A x(t) + B u_s(t) + w(t) \label{eqn:appC:ss1} \\
y(t) =& C x(t) + D u_s(t) + v(t) \label{eqn:appC:ss2}
\end{align}
where (1) $x(t)$, $u_s(t)$ and $y(t)$ are, respectively, the state, input and output vectors; (2) $w(t)$ is the state noise vector and $v(t)$ is the measurement noise vector, where both vectors are zero-mean Gaussian white noise; and (3) $(A,B,C,D)$ is a state space quadruple. We assume the dimensions of the vectors and matrices are compatible. The subscript $s$ in $u_s(t)$ plays the same role as the Symbol $s$ in the main text. We can use different $s$ to choose different input $u_s(t)$. We can determine the mean state vector ${\mathbf E}[x(t)]$ by taking expectation on both sides of Equation \eqref{eqn:appC:ss1} to obtain: 
\begin{align}
\frac{d{\mathbf E}[x(t)]}{dt} =& A {\mathbf E}[x(t)] + B u_s(t) 
\end{align}
Note that ${\mathbf E}[x(t)]$ plays the same role as $\sigma_s(t) = {\mathbf E}[n_R(t) | s]$. 

The filtering problem for a LTI system is to estimate the state vector $x(t)$ from the continuous history of the output $y(t)$. A method to realise filtering is to use an observer \cite{Franklin:cont}:
\begin{align}
\frac{d\hat{x}(t)}{dt} =& A \hat{x}(t) + B u_s(t) + K (\hat{y}(t)-y(t))  \label{eqn:appC:obs:ss1}  \\
\hat{y}(t) =& C \hat{x}(t) + D u_s(t)   \label{eqn:appC:obs:ss2} 
\end{align}
where $K$ is the observer gain matrix. The vector $\hat{x}(t)$ is the estimated state vector from the past history of the output $y(t)$. The expectation of $\hat{x}(t)$, i.e. ${\mathbf E}[\hat{x}(t)]$, plays the same role as ${\mathbf E}[n_R(t) | s, {\cal B}(t)]$. 

We are interested to study the difference $e(t) = {\mathbf E}[\hat{x}(t)] - {\mathbf E}[x(t)]$. By using Equations \eqref{eqn:appC:ss1}, \eqref{eqn:appC:ss2}, \eqref{eqn:appC:obs:ss1} and \eqref{eqn:appC:obs:ss2}, it can be shown that $e(t)$ obeys the differential equation
\begin{align}
\frac{de(t)}{dt} =& (A+KC) e(t) 
\end{align}
This means that if the matrix $A+KC$ is asymptotically stable, then the difference $e(t)$ tends to zero for sufficiently large $t$. This result suggests that the difference between ${\mathbf E}[n_R(t) | s, {\cal B}(t)]$ and $\sigma_s(t)$ can be small if the filtering error is stable. 

%
%
%



\newpage
\listoffigures 
\newpage

%
%

%
%

\end{document}